\documentclass[final,1p,times]{elsarticle}

\usepackage[T1]{fontenc}
\usepackage{graphicx}
\usepackage{amssymb}
\usepackage{amsmath}
\usepackage{lipsum}
\usepackage{multirow}
\usepackage{booktabs}
\usepackage{comment}
\usepackage{float}
\usepackage{hyperref} 
\usepackage{subfigure}  
\usepackage[table]{xcolor} 
\usepackage[linesnumbered,ruled,vlined]{algorithm2e}
\graphicspath{{Figure/}} 

\journal{arXiv}

\begin{document}

\begin{frontmatter}



\title{DroidTTP: Mapping Android Applications with TTP for Cyber Threat Intelligence}
\author[inst1]{Dincy R Arikkat}   
\author[inst2,inst1]{Vinod P.}
\author[inst1]{Rafidha Rehiman K. A.}
\author[inst4]{Serena Nicolazzo}
\author[inst3]{Marco Arazzi}
\author[inst3]{Antonino Nocera}
\author[inst2]{Mauro Conti}

\affiliation[inst1]{organization={Department of Computer Applications},
            addressline={Cochin University of Science and Technology}, 
            state={Kerala},
            country={India}}        
\affiliation[inst2]{organization={Department of Mathematics},
            addressline={University of Padua}, 
            state={Padua},
            country={Italy}}
\affiliation[inst3]{organization={Department of Electrical, Computer and Biomedical Engineering}, 
            addressline={University of Pavia}, 
            state={Pavia},
            country={Italy}}
\affiliation[inst4]{organization={Department of Computer Science},
            addressline={University of Milan}, 
            state={Milan},
            country={Italy}}

\begin{abstract}

The rapid expansion of the Internet of Things (IoT) and mobile technologies has led to an increased reliance on Android devices for sensitive operations such as banking, online shopping, and communication. While Android remains the dominant mobile operating system, its widespread adoption has made it a prime target for cyber threats, particularly Advanced Persistent Threats (APT) and sophisticated malware attacks. Traditional malware detection methods focus primarily on binary classification, failing to provide insights into the Tactics, Techniques, and Procedures (TTPs) used by adversaries. Understanding how malware operates is essential for strengthening cybersecurity defenses.

To bridge this gap, we present DroidTTP, a solution designed to map Android malware behaviors to TTPs as defined by the MITRE ATT\&CK framework. This system empowers security analysts with deeper insights into attacker methodologies, enabling more effective defense strategies. In this work, we curated a novel dataset explicitly designed to link MITRE TTPs to Android applications. Moreover, we developed an automated solution leveraging the Problem Transformation Approach (PTA) and Large Language Models (LLMs) to map Android applications to both Tactics and Techniques. Furthermore, we exploited LLMs for TTP predictions and experimented with two different strategies, specifically Retrieval-Augmented Generation with prompt engineering and LLM fine-tuning. Our approach follows a structured pipeline, including dataset creation for Android TTP prediction, hyperparameter tuning,  data augmentation, feature selection, model development for prediction, and interpreting the model decision using SHAP. For Tactic classification, the Llama model achieved the highest performance among LLMs, with a Jaccard Similarity score of 0.9583 and a Hamming Loss of 0.0182. Similarly, for Technique classification, Llama outperformed other LLMs, attaining a Jaccard Similarity score of 0.9348 and a Hamming Loss of 0.0127. However, the Label Powerset XGBoost model demonstrated superior performance, achieving a Jaccard Similarity score of 0.9893 and a Hamming Loss of 0.0054 for Tactic classification, while attaining a Jaccard Similarity score of 0.9753 and a Hamming Loss of 0.0050 for Technique classification. Although XGBoost slightly outperformed the fine-tuned LLMs, the performance gap remains narrow, highlighting the potential of LLM-based approaches in TTP classification.
\end{abstract}




\end{frontmatter}


\section{Introduction}
\label{introduction}
In recent years, the digital landscape has undergone significant transformations, with critical assets such as financial data and personal information increasingly shifting to online platforms \cite{van2022digital}. The ubiquity of mobile and IoT devices in daily life has primarily driven this shift. Unlike traditional Personal Computers (PCs), these devices not only facilitate daily tasks but also handle a vast amount of sensitive data essential for activities ranging from banking \cite{wu2021android} to online shopping and entertainment \cite{bakour2021visdroid}. By 2023, Android had solidified its position as the world's leading mobile operating system, capturing over $70$\% of the global market share and surpassing its competitors \footnote{\url{https://www.statista.com/statistics/921152}}. Its popularity comes from its open-source, cost-effective, and user-friendly nature \cite{islam2023android}, factors that have contributed to the proliferation of over $2.61$ billion applications\footnote{\url{https://www.businessofapps.com/data/google-play-statistics}} available through the Google Play Store.

\par However, this widespread adoption has also made Android devices attractive targets for cybercriminals, particularly Advanced Persistent Threat (APT) groups \cite{zulkefli2020sentient}. The escalation of smartphones has allowed users to make mobile payments and store sensitive information such as login credentials, increasing the risks for both individuals and organizations. As a result, the past decade has seen a significant rise in mobile attacks, introducing a range of sophisticated threats that specifically target mobile platforms \cite{goyal2010literature}. The nature of threats emerging from mobile devices often parallels those affecting traditional PCs. These threats are not limited to conventional malware like worms, Trojans, and viruses; they extend to sophisticated cyber attacks that can compromise security and privacy or even gain complete control over a device. Also, mobile attacks spread malicious content rapidly through technologies such as 5G and Wi-Fi, which provide seamless Internet connectivity. This connectivity renders mobile devices particularly attractive to cyber criminals, who can deploy a variety of attack vectors, from exploiting mobile sensors to executing application-level malware. As such, understanding and mitigating these attacks is crucial in safeguarding both personal and organizational assets from the diverse array of threats targeting mobile platforms.

\par Existing studies mainly focused on mobile malware detection and classification to address various security challenges \cite{yan2018survey,vinayakumar2017deep}. However, there is a notable lack of research on cyber attribution, which aims to track attack Tactics and Techniques employed by an attacker. Current Machine Learning (ML) methods typically offer classification for mobile users and app security analysts, indicating whether an application is likely to be malicious or benign. This approach falls short of fully addressing the complexities of malware detection. For example, suppose that a mobile application is found to exhibit suspicious behavior, such as accessing sensitive user data without permission. Simply labeling this app as malicious or belonging to a malicious family, does not explain how it conducts its attack. Understanding the specific Tactics, like exploiting a vulnerability in the operating system or using social engineering to trick users, can provide deeper insights into the threat and inform better defensive strategies. Mapping Tactics, Techniques, and Procedures (TTPs) to malware behavior would empower security professionals to not only identify threats but also understand the attacker's motives and methods. 

\par The MITRE ATT\&CK \footnote{\url{https://attack.mitre.org/}} framework provides a structured overview of TTPs associated with attacks. It serves as a valuable resource for security professionals, providing a common language and detailed documentation that helps to understand attacker behavior. By categorizing TTPs, the framework enables security analysts to better identify, analyze, and respond to threats in a systematic way. However, there is currently a lack of automated methods to correlate Android applications with particular TTPs. As mobile malware becomes increasingly sophisticated, attackers are constantly evolving their Tactics, making it essential for security analysts to have tools that can automatically map application behaviors to TTPs. 
To bridge this gap, we propose \textit{DroidTTP}, a system designed to identify the TTPs associated with Android malware. 

To build the DroidTTP framework, we have created a new dataset to train automated models that map TTPs to Android applications. Subsequently, we implemented a multilabel Tactic and Technique classification model to identify the TTPs associated with these applications (apps). Our methodology leverages Problem Transformation Approaches (PTA), particularly useful in multi-label classification, and Machine Learning Techniques to enhance TTP prediction. Beyond traditional models, we explore the potential of Large Language Models (LLMs) by implementing Retrieval-Augmented Generation (RAG) with prompt engineering for dynamic and context-aware predictions. In particular, we aimed to evaluate the effectiveness of LLMs in Android malware detection by analyzing features extracted from application components and mapping them to attacker behaviors. This approach seeks to determine whether LLMs can infer malicious intent based on patterns of permissions, activities, receivers, services, and inter-component communications, thereby enhancing behavioral analysis in malware classification. Additionally, we fine-tuned an LLM to evaluate its effectiveness in predicting TTPs based on Android application features.

The main contributions of our work are as follows.
\begin{itemize}
    \item We introduce a novel dataset designed explicitly for mapping MITRE TTPs to Android applications. This dataset enables researchers and security professionals to analyze the behavior of various Android apps in relation to established cybersecurity frameworks.
    
    \item We propose an enhanced feature selection strategy for multi-label classification that builds upon the \textit{SelectKBest}\footnote{\url{https://scikit-learn.org/stable/modules/generated/sklearn.feature_selection.SelectKBest.html}} method, a commonly used univariate feature selection method that identifies the top \( K \) features by applying statistical tests like chi-square or ANOVA F-value. Our approach involves a two-step process: first, performing label-specific feature selection to capture the most relevant features for each label, and second, generalizing this selection across multiple datasets to identify robust features.
    
    \item  We develop an automated system that maps Android applications to different MITRE Tactic labels. This system utilizes advanced multi-label classification algorithms to analyze app behavior and associate it with specific Tactics. 
    
    \item In addition to Tactic mapping, we implement an automated system that assigns Techniques from the MITRE framework to Android applications. 
    \item We investigate the potential of LLMs for predicting Tactics and Techniques in both Retrieval-Augmented Generation (RAG) and fine-tuning settings. Our study evaluates various LLMs, explores different prompt engineering strategies, and analyzes their effectiveness in enhancing TTP prediction.

\end{itemize}

The remainder of this paper is organized as follows. Section \ref{sec:related_work} discusses related works. Section \ref{sec:methodology} explains the methodology we used to develop DroidTTP. Section \ref{sec:result} presents the experimental evaluation and discussion of our findings. Finally, Section \ref{sec:conclusion} concludes with information and outlines future research directions.

\section{Literature Review}
\label{sec:related_work}
Recently, numerous studies have focused on Android malware detection, leveraging various techniques such as static features, dynamic features, hybrid features, and visualization analysis \cite{qiu2020survey}. Kim et al. \cite{kim2018multimodal} proposed an Android malware detection framework based on static features to capture multiple aspects of application properties. Their framework comprehensively examines Android Package Kit (APK) files by extracting seven distinct feature types: Environmental, Component, Permission, Shared Library Function Opcode, Method API, Method Opcode, and String data. They used a multimodal Deep Learning (DL) architecture where separate networks process different feature types independently before combining their outputs into a final network. The MADAM \cite{saracino2016madam} employs a multi-level analysis strategy, examining features across the package, user, application, and kernel levels. It combines two parallel classifiers with a behavioral signature-based detector to strengthen its malware identification capabilities.
In \cite{arora2019permpair}, Arora et al. introduced PermPair, a model that detects malware by analyzing relationships between permission pairs found in the Android application manifest, constructing graph representations. Their approach demonstrated strong performance with $95.44\%$ detection accuracy, which exceeded both state-of-the-art detection methods and commercial anti-malware solutions. Building on graph-based approaches, the GHGDroid \cite{shen2024ghgdroid} framework takes a broader perspective by creating comprehensive heterogeneous graphs that map relationships between Android applications and their interactions with sensitive APIs. By employing multi-layer graph convolution networks, their approach achieved an F1 score of $99.17\%$. 

Despite advancements in malware detection, the identification of attack behaviors such as Tactics and Techniques remains an underexplored area in the current literature. While some studies, including TTPDrill \cite{husari2017ttpdrill} and rcATT \cite{legoy2020automated}, focus on extracting TTPs from threat reports, their reliance on term frequency-based methods limits the accuracy and completeness of TTP identification. These methods fail to capture the contextual relationships within CTI reports, thereby limiting their effectiveness. Shin et al. \cite{shin2023exploiting} propose a TTP extraction approach based on GloVe embedding, which uses a co-occurrence matrix to capture semantic relationships. Their results show a correlation of up to $0.96$ between the co-occurrence matrix and the embedding performance for each Tactic.

Recent research also investigates mapping TTPs to network traffic. In \cite{sharma2023ttp,sharma2023radar}, Sharma et al. introduced an ML-based approach for malware detection in network traffic by utilizing adversarial behavior knowledge represented as TTP. They extracted TTP features from network traffic and integrated them into ML models, demonstrating superior performance compared to state-of-the-art malware detection methods.
In the Android security domain, Fairbanks et al. \cite{fairbanks2021identifying} identified attack Tactics by analyzing control flow graphs in malware samples.  Their methodology employed the Inferential SIR-GN model for node representation and utilized Random Forest classification, with SHAP analysis providing insight into feature importance and subgraph relevance. This approach achieved a $92.7\%$ F1 score in Tactic detection. However, the scope of their research was limited to seven specific Tactics and did not extend to the broader range of attack Techniques used by malicious actors. Despite these advances, there has not been a comprehensive study that maps both attack Tactics and Techniques to Android applications, and no dataset exists that captures these relationships in detail.

Our research addresses these critical gaps through several key innovations. First, we compiled a benchmark dataset that captures multiple Tactics and Techniques related to Android applications. This dataset serves as a foundation for our multi-label classification system, which maps applications to their corresponding MITRE Tactics and Techniques. This work also introduces an adaptive feature selection framework that optimizes feature sets based on the characteristics of different training datasets, leading to enhanced model performance and broader applicability. To the best of our knowledge, while LLMs have been widely applied in various cybersecurity tasks \cite{chen2025aecr,zhang2025attackg+,shafee2024evaluation,hu2024llm}, no existing study has specifically explored their use in mapping Android applications to TTPs. In this work, we investigate the potential of LLMs by leveraging RAG and fine-tuning Techniques for TTP prediction.

\section{Background}
\label{sec:background}
In this section, we provide the necessary background information for our study, covering Tactics, Techniques, and Procedures, Adversarial Tactics, Techniques, Common Knowledge (ATT\&CK), and Retrieval-Augmented Generation.

Table \ref{tab:SystemSymbols} summarizes the acronyms used in this paper.

\begin{table}
\footnotesize
\centering
  \caption{Summary of the acronyms used in the paper}
  \begin{tabular}{ll}
\hline
    \textbf{Symbol} & \textbf{Description}\\
\hline
    ANN & Approximate Nearest Neighbor\\
    APK & Android Package Kit\\
    APT & Advanced Persistent Threat\\
    DL & Deep Learning\\
    DT & Decision Trees\\
    ICS & Industrial Control System\\
    IoC & Indicator of Compromise\\
    IoT & Internet of Things\\
    LLM & Large Language Model\\
    ML & Machine Learning\\
    MLP & Multi-Layer Perceptron\\
    PC & Personal Computer\\
    PTA & Problem Transformation Approach\\
    RAG & Retrieval-Augmented Generation\\
    RF & Random Forest\\
    SHAP & SHapley Additive exPlanations\\
    SLM & Small Language Model\\
    TTPs & Tactics, Techniques, and Procedures\\
\hline
\end{tabular}
\label{tab:SystemSymbols}
\end{table}

\subsection{Tactics, Techniques, and Procedures}
In cybersecurity, understanding the methodologies, intentions, and actions of adversaries is fundamental to building an effective defense framework. An effective way to analyze and classify adversarial activities is by understanding TTPs. TTPs describe the behavior of attackers that provides both a strategic overview and tactical insights into their operations. 
Although Indicators of Compromise (IoCs) such as IP addresses, file hashes, or domain names are invaluable for real-time threat detection, they typically lack the context necessary to interpret the intentions and behavior of attackers. TTPs, on the other hand, serve as higher-level IoCs that provide a more comprehensive view of how an attack unfolds. 
The three components of TTPs are:
\begin{itemize}
    \item \textit{Tactics:}  Represents the high-level \textit{goals or objectives} that an adversary aims to achieve during an attack. 
    In the context of Android security, these objectives could range from data theft and system infiltration to denial of service. 
    \item \textit{Techniques:} Describe how the attacker will perform the Tactic. They are the specific \textit{methods or actions} used to achieve the objective, providing more granular insights into the attacker’s approach.  In Android security, several Techniques can be employed to execute a particular Tactic. For example, if the attacker’s Tactic is to gain access to the system, a common Technique could be \textit{phishing}. 
    \item \textit{Procedures:}  Procedures are the exact \textit{steps or actions} 
    that provide a step-by-step description of how the attacker operationalizes their methods to achieve their objectives. For example, consider the procedure steps where the adversary utilizes phishing as a Technique: (i) The attacker crafts a phishing email disguised as an urgent message from the victim’s Android banking app, stating that the account has been locked and needs to be reset. (ii) The email contains a link to a fake login page, which is designed to look identical to the official banking app login screen. (iii) The victim, believing the message is legitimate, clicks the link and enters their login credentials into the fake page. (iv) The attacker then captures the credentials and uses them to access the victim’s bank account, potentially conducting fraudulent transactions. 
\end{itemize}

\subsection{Adversarial Tactics, Techniques, and Common Knowledge}
The MITRE Adversarial Tactics, Techniques, and Common Knowledge (ATT\&CK) framework is a comprehensive knowledge base that catalogs adversary behaviors. Developed by MITRE, this framework provides a structured approach to understanding how attackers operate. It plays a vital role in threat intelligence, incident response, red teaming, and adversary emulation by mapping detected behaviors to known attack Techniques. The ATT\&CK framework is organized into multiple matrices, each tailored to specific environments and attack vectors. These include the \textit{Enterprise}, \textit{Mobile} (Android and iOS), and \textit{Industrial Control Systems (ICS)} matrices, which help analysts focus on different operating systems or environments. 

Each matrix is further broken down into Tactics (e.g., Initial Access, Execution, Exfiltration, etc.). Techniques are then listed under each Tactic, providing detailed descriptions of the methods adversaries may employ to achieve their goals (e.g., phishing, credential dumping, or lateral movement). For example, consider the \texttt{Initial Access 
 (TA0027)} Tactic in an Android attack scenario. A common technique used by attackers to achieve this goal is \texttt{Phishing (T1660)}\footnote{\url{https://attack.mitre.org/techniques/T1660}}, where social engineering is used to trick victims into giving the attacker access to their devices. The attack progresses through a sequence of meticulously planned steps, further explained in the procedure examples and referenced within the specific Technique page.

Moreover, Techniques can have sub-techniques 
which further details specific variations of the Technique. For example, the \texttt{Input Capture (T1417)}\footnote{\url{https://attack.mitre.org/techniques/T141}} technique focuses on how adversaries intercept user input on a mobile device. This Technique is divided into sub-techniques that outline distinct methods of input interception: \texttt{Keylogging (T1417.001)}, 
and \texttt{GUI Input Capture (T1417.002)}. 
 A single Technique can belong to multiple Tactics. For example, the Input Capture Technique is used in both Credential Access and Collection Tactics.

\subsection{Retrieval-Augmented Generation}
Retrieval-Augmented Generation (RAG) \cite{lewis2020retrieval} represents a fusion of information retrieval and language generation technologies. This approach enhances AI language models by connecting them with external knowledge sources that enable more informed and accurate responses. Unlike traditional language models that rely solely on their training data, RAG actively draws upon current information when generating responses.

\begin{figure}
    \centering
    \includegraphics[width=\linewidth]{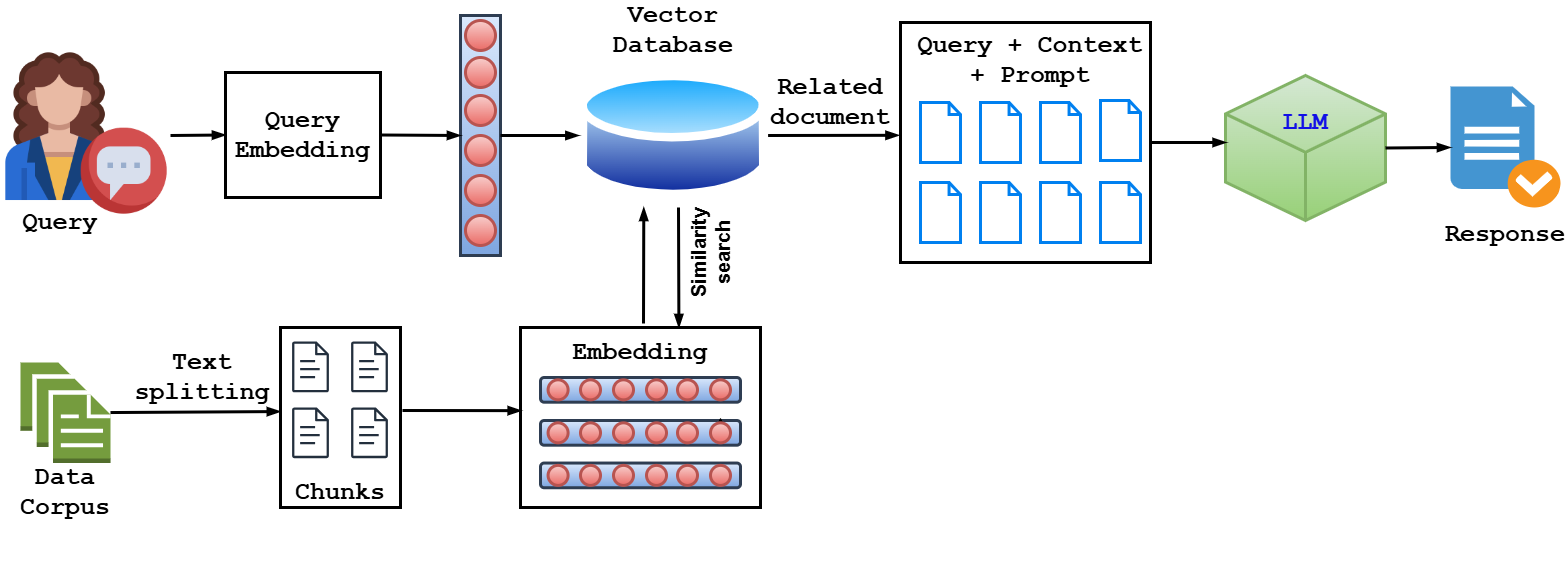}
    \caption{Architecture of Retrieval-Augmented Generation }
    \label{fig:rag_arc}
\end{figure}

The RAG workflow is illustrated in Figure \ref{fig:rag_arc}. At its core, RAG functions through a two-stage process. The first stage involves information \textit{retrieval}, where the system searches through external knowledge sources to find content relevant to the current query. This works by converting both the user's question and the available reference documents into mathematical representations called embeddings. 
It then scans similarly encoded documents in its knowledge base, using mathematical comparison of embeddings to identify the most semantically relevant information.

The second stage leverages an LLM's \textit{generative} capabilities. Rather than relying solely on its pre-trained knowledge, the model receives both the user's query and the retrieved relevant information (context) from the vector database. This allows the LLM to craft responses that incorporate specific, factual details from the retrieved sources while maintaining natural language fluency. The result is more accurate and contextually appropriate than what could be achieved by either retrieval or generation alone. This combination of dynamic knowledge access and sophisticated language generation represents a significant leap forward in AI’s ability to provide precise, contextually enriched responses.

\begin{figure}[!ht]
    \centering
    \includegraphics[width=\linewidth]{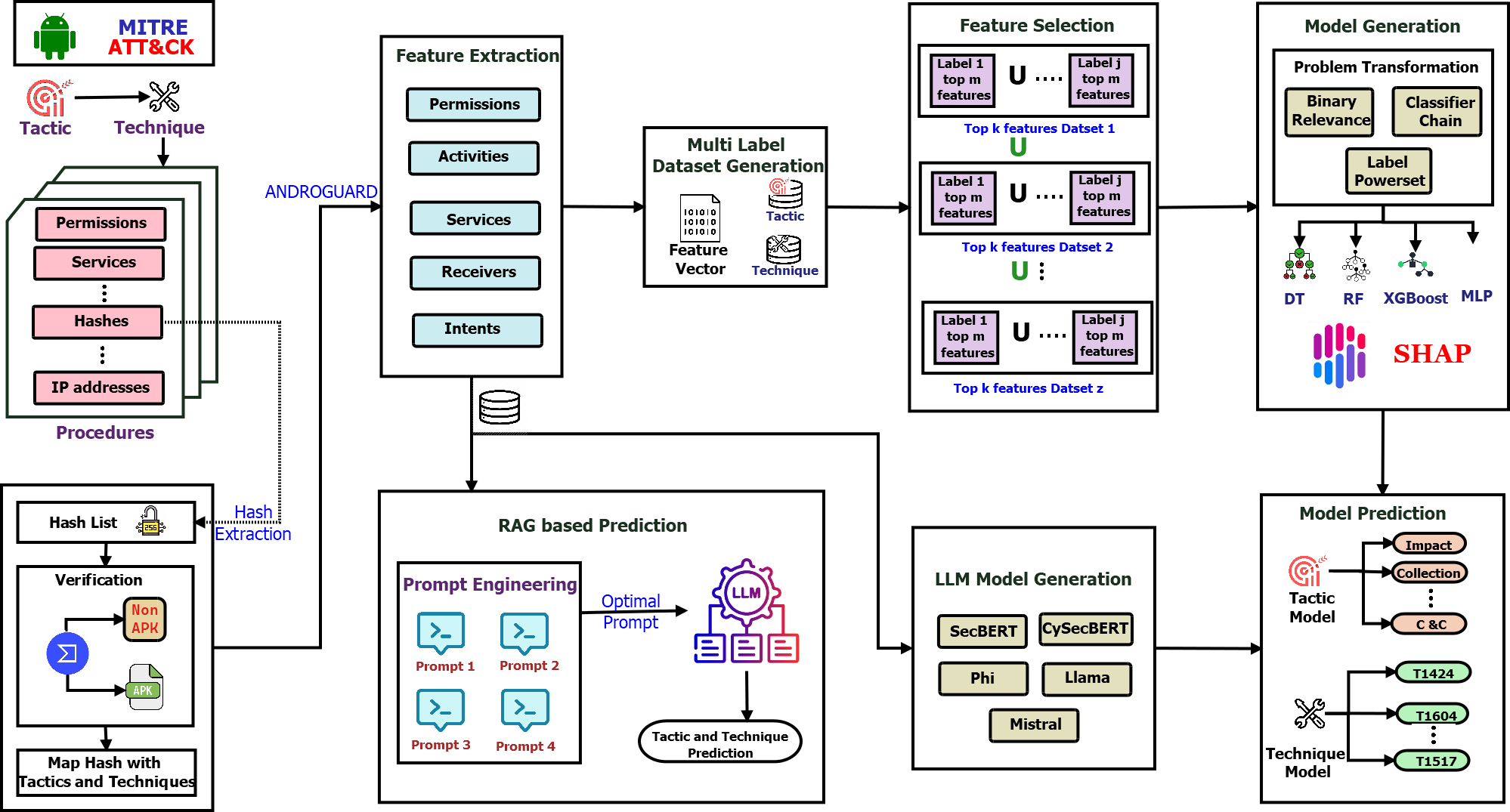}
    \caption{Architecture of Android Tactic and Technique Prediction Model}
    \label{fig:architecture}
\end{figure}
\section{Data Collection and Processing}
\label{sec:methodology}
In this section, we present the first steps of the proposed methodology used in this study. The architecture of the proposed system is illustrated in Figure \ref{fig:architecture}. The initial phase involves collecting data for experimentation. Following data collection, key features are identified using a feature selection mechanism. 
\subsection{Data Collection}
\label{sub:collection}
In this phase, we gather data to generate a model that maps Android applications to MITRE Tactics and Techniques. To the best of our knowledge, no existing dataset provides MITRE Tactic and Technique classification for Android apps. Therefore, we initiated the data collection process by compiling hashes of Android applications. This process involved manually analyzing the Tactic and 
Technique information outlined in the MITRE ATT\&CK Mobile Android Matrix \footnote{\url{https://attack.mitre.org/matrices/mobile/android}}. 

As outlined in Section \ref{sec:background}, the MITRE framework categorizes adversarial behavior into distinct, high-level Tactics, each of which encompasses multiple specific Techniques. The Android matrix covers Tactics, including \textit{Initial Access, Execution, Persistence, Privilege Escalation, Defense Evasion, Credential Access, Discovery, Lateral Movement, Collection, Command and Control, Exfiltration,} and \textit{Impact}. Moreover, each Technique is accompanied by technical procedures, and we used these procedure details to collect the Android applications. The data collection process involves the following three steps: 
\begin{figure}[ht]
    \centering
    \includegraphics[width=0.8\linewidth]{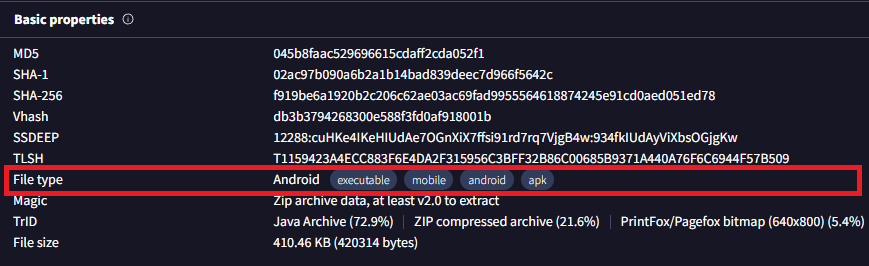}
    \caption{VirusTotal report confirming the hash matches an Android app.}
    \label{fig:virustotal_hash}
\end{figure}
\begin{enumerate}
    \item \textbf{Procedure References Analysis.} The first step involved a detailed examination of the procedure references associated with each Technique in the MITRE ATT\&CK Mobile Android Matrix. These references, contributed by security practitioners, outline how attackers exploit specific vulnerabilities to carry out malicious activities on Android devices. They provide critical technical details such as permissions requested by the applications, inter-process communication intents, app activities, and additional IoCs, such as suspicious URLs or IP addresses. For example, the \texttt{Audio Capture (T1429)}\footnote{\url{https://attack.mitre.org/techniques/T1429}} Technique, under the \texttt{Collection} Tactic, is linked to $49$ references detailing scenarios in which malware records audio without user consent, abuses sensitive permissions, or exfiltrates captured data to external servers. Many of these references also provide cryptographic hashes (e.g., SHA1) of malicious files. For instance, one reference from Lookout\footnote{\url{https://www.lookout.com/threat-intelligence/article/iranian-spyware-bouldspy}} provides SHA1 hashes of \textit{BouldSpy} samples along with their associated Command and Control details. By systematically analyzing these procedure references, we can compile a set of hashes. We manually analyzed each reference link provided in the technical details and collected the corresponding hashes. The number of hashes collected for this study is detailed in Section \ref{subsec:dataset}.

\item \textbf{Verification Using VirusTotal.} The next step involved verifying the collected hashes using VirusTotal\footnote{\url{https://www.virustotal.com/gui/home/search}}, a widely recognized threat intelligence platform that provides comprehensive security reports on files, URLs, and IPs. VirusTotal aggregates data from multiple antivirus engines and other security tools, making it a powerful platform for threat analysis. It scans each hash to determine the file type (e.g., APK, Windows, etc.) and assesses whether it is malicious. For example, we verified a hash extracted from a Lookout Threat Intelligence reference on BouldSpy spyware using VirusTotal\footnote{\url{https://www.virustotal.com/gui/file/f919be6a1920b2c206c62ae03ac69fad9955564618874245e91cd0aed051ed78/details}}. The report confirmed that the hash was compatible with an Android app, as shown in Figure \ref{fig:virustotal_hash}. Using VirusTotal, we eliminated irrelevant data, such as hashes that do not correspond to Android applications.

\item \textbf{Labeling Techniques and Tactics.} The final step of the data collection process involved labeling each verified hash with one or more MITRE ATT\&CK Techniques and their corresponding Tactics. This labeling is based on the procedure references analyzed in the first phase. For each verified hash, we identified the associated Technique(s) and Tactic(s) it represented. For example, suppose that a hash is included in the procedure reference for the \textit{Audio Capture} Technique. We label it with the Audio Capture Technique ID and associate it with the corresponding Tactic, which in this case is \textit{Collection}.
Moreover, since an app can exhibit multiple behaviors corresponding to different Techniques and Tactics, we used a multi-labeling approach. For instance, an app exploiting the \textit{Call Control} Technique could be labeled under the \textit{Collection} Tactic (due to data gathering), \textit{Command and Control} Tactic (for communication with a C2 server), and \textit{Impact} Tactic (because of potential disruption to user communication). We applied this approach to label each app based on all the relevant Techniques and Tactics it exhibited. As a result, we created a dataset that includes Android app hashes along with their associated Technique IDs and corresponding Tactics.
\end{enumerate}

\subsection{Feature Extraction}
After collecting the hashes of Android apps, the next step involves extracting relevant features for a supervised learning classification task. In this study, we focus on extracting Android app features that are commonly referenced in the MITRE procedure details. For instance, the procedure reference for the \texttt{T1398 - Boot or Logon Initialization Scripts} Technique under the \texttt{Persistence} Tactic\footnote{\url{https://thehackernews.com/2014/01/first-widely-distributed-android.html}} examines the permissions requested by the malware, providing insight into its persistence mechanisms.
\textit{Android.Oldboot bootkit}. Similarly, static features such as Activities, Intents, Services, Receivers, Strings, etc., are also detailed in the procedure reference documents. Based on the frequency of these static features mentioned in the references, we extracted the following features. 
\begin{itemize}
    \item \textit{Permissions}: These are access rights requested by the app to perform specific actions. For example, navigation apps and social media platforms often request \texttt{ACCESS\_FINE\_LOCATION} permission to access the device's GPS location for location-based features. 
    \item \textit{Activities}: Activities represent individual screens or interactions within the app's user interface. Each activity defines a specific user experience. A login screen (\texttt{LoginActivity}) enables users to enter their credentials for authentication. 
    \item \textit{Services}: These are background processes that operate independently of the user interface. Services can handle long-running tasks, such as playing music or fetching data from the Internet. For example, the \texttt{MusicPlayerService} in music streaming apps like Spotify plays music in the background, even when the app is minimized or the screen is off. 
    \item \textit{Receivers}: Receivers are Android components designed to listen for system events or inter-application broadcasts, enabling applications to dynamically respond to environmental changes. They facilitate event-driven execution, allowing apps to react to network state transitions, incoming messages, or system-level triggers. For instance, \texttt{BatteryReceiver} detects battery level fluctuations, enabling applications to implement power-saving mechanisms or monitor energy consumption efficiently. 
    \item \textit{Intents}: Intents facilitate communication between application components and the system, enabling seamless interaction and execution of predefined actions. Intent actions define the specific operations an application can perform, while intent categories provide contextual information to refine the behavior. For example, the `\texttt{ACTION\_VIEW}` intent with a URL parameter triggers a web browser to open a webpage. Malware often exploits this mechanism to redirect users to malicious websites, leveraging intent-based redirection as a vector for phishing, drive-by downloads, or credential theft. 
\end{itemize}
To extract the features, we performed a static analysis. Static analysis examines the application's code without executing it, thus avoiding the risks associated with running potentially malicious software.
Static analysis typically involves reverse engineering Android Package (APK) files, which are compressed archives containing essential components such as the manifest file and other resources. The \textit{AndroidManifest.xml} file within the APK is a key component for static analysis, as it contains valuable information about the app's structure and capabilities. 
To automate the feature extraction process, we used VirusTotal, which provides static details about Android APKs through Androguard\footnote{\url{https://github.com/androguard/androguard}}. Androguard offers comprehensive information about the app's permissions, activities, services, etc. To collect features, we first created a VirusTotal account and developed a script to query the platform for each app's hash. VirusTotal responds with data in JSON format, which our script processes to extract static features.  Finally, we converted these features into binary feature vectors suitable for ML algorithms.

We denote this feature vector as \( \mathcal{F} = (f_1, f_2, \ldots, f_s) \), where \( s \) is the number of unique static features. These unique features are determined by consolidating all features from the various hashes and identifying the distinct ones.  For each specific hash, the features are encoded as:

\[ f_i = \begin{cases} 
1 & \text{if the i-th feature is present} \\
0 & \text{otherwise}
\end{cases} \]

\subsection{Feature Selection}
This phase addresses the crucial ML aspect of dimensionality reduction. High-dimensional datasets can lead to overfitting, in which the model performs well on the training data but struggles with unseen data. In addition, processing a large number of features can be computationally expensive. Thus, effective feature selection is essential to improve the accuracy of the model.

\textit{SelectKBest} is a widely used \cite{yoga975hybrid,al2024machine} feature selection method that identifies the top $k$ features based on their statistical significance in relation to the target variable. This approach evaluates each feature individually and ranks attributes according to a chosen statistical test, such as the chi-squared test or ANOVA F-value. Although this approach is effective in many cases, however, it has limitations in multi-label classification tasks. Specifically, features relevant to one label may be overlooked when evaluated independently, as this approach does not account for inter-dependencies among labels. Moreover, feature selection is often carried out on a specific training set determined by a random seed. However, this practice can lead to significant variability in results. Features selected from one dataset may not perform optimally when applied to a different dataset generated with a different random seed. This inconsistency arises because the selected features are solely based on the characteristics of a single training dataset, which may not be representative of other datasets. To address these challenges, we propose a novel feature selection strategy that leverages \textit{SelectKBest} but enhances it for multi-label classification tasks. The proposed mechanism follows a two-step process that ensures consistent feature selection while accounting for label interdependencies to improve classification accuracy, namely:

\begin{enumerate}
    \item \textbf{Label Specific Feature Selection.} The first step focuses on identifying features that are most relevant to each label within a specific dataset. To achieve this, we used the chi-square test in combination with the SelectKBest method, which ranks top $m$ features according to their association with the target label. To calculate the chi-square value, we computed the observed frequency and the expected frequency. The Observed Frequency \(\mathbf{O_{ij}}\) quantifies the number of occurrences in which feature \( f_i \) appears alongside label \( y_j \) in the dataset. In contrast, the Expected Frequency \(\mathbf{E_{ij}}\) represents the anticipated occurrence of \( f_i \) given \( y_j \), assuming that there is no statistical dependency between them. Expected Frequency ($E_{ij}$) is calculated by the Equation \ref{eq:EF}:
\begin{equation}
E_{ij} = \frac{T_{f_i} \times T_{y_j}}{N}
\label{eq:EF}
\end{equation}
Where \( T_{f_i} \) is the total occurrences of feature \( f_i \),  \( T_{y_j} \) is the total occurrences of label \( y_j \), and \( N \) is the total number of instances in the dataset.
The chi-square statistic is then computed for each feature-label pair using the following Equation \ref{eq:chi}:
\begin{equation}
\chi^2 = \sum_{i,j} \frac{(O_{ij} - E_{ij})^2}{E_{ij}}
    \label{eq:chi}
\end{equation}

After calculating the chi-squared statistic for each feature-label pair, the \textit{SelectKBest} method ranks the features based on their chi-squared values. Features with higher chi-squared scores are considered more statistically significant in their relationship with the label and are therefore selected for further use. For each label, the top $m$ features with the highest chi-squared scores are selected. 

\item \textbf{Generalizable Feature Subset.} In the second step, we apply the same procedure as in the first step to \(z\) different datasets, each generated using a different random seed. We analyze each dataset separately to identify the most important features. Finally, we combine the top features from all \(z\) datasets and extract only the unique ones. The comprehensive process of feature selection is presented in Algorithm 1.

\end{enumerate}
\begin{algorithm}
\caption{Feature Selection Process}
\KwIn{$\{\mathcal{D}_1, \mathcal{D}_2, \ldots, \mathcal{D}_{z}\}$: $z$ datasets generated by different random seeds, $\mathcal{L}$: Set of labels, $m$: Number of top features to select for each label}
\KwOut{$\mathcal{F}_{\mathcal{G}}$: Generalizable subset of unique top features across all $z$ datasets}

Initialize $\mathcal{F}_{\mathcal{G}} \gets \emptyset$ 

\For{$i \gets 1$ \KwTo $z$}{
    Initialize $\mathcal{F}_{\mathcal{L}}^{(i)} \gets \emptyset$ 
    
    \For{each label $l \in \mathcal{L}$}{
        \For{each feature $f_j$ in $\mathcal{D}_i$}{
            
            $\chi^2 = \sum_{i,j} \frac{(O_{ij} - E_{ij})^2}{E_{ij}}$\;
            
            Store $\chi^2$ score for feature $f_j$\;
        }
        
        Rank features $f_j$ by their $\chi^2$ score in descending order\;
        Select top $m$ features $f_1, f_2, \ldots, f_m$ from ranked list\;
        $\mathcal{F}_{\mathcal{L}}^{(i)}$= $\mathcal{F}_{\mathcal{L}}^{(i)}$ + $\{f_1, f_2, \ldots, f_m\}$\;
    }
    
   $\mathcal{F}_{\mathcal{L}}^{(i)}$ = set($\mathcal{F}_{\mathcal{L}}^{(i)}$) \;
   
   $\mathcal{F}_{\mathcal{G}}$ = $\mathcal{F}_{\mathcal{G}}$ +$\mathcal{F}_{\mathcal{L}}^{(i)}$\;
}

$\mathcal{F}_{\mathcal{G}}$ = set( $\mathcal{F}_{\mathcal{G}}$)\;
return $\mathcal{F}_{\mathcal{G}}$ 
\label{algo:feature}
\end{algorithm}

\section{Classification Model Generation}
This section is devoted to presenting the second phase of our methodology in Figure~\ref{fig:rag_arc}.
In this phase, we develop a classification model for identifying Tactics and Techniques using the collected data. To achieve this, we used the Problem Transformation Approach.

\subsection{Problem Transformation Approach}
Traditional supervised learning typically involves single-label models, where each instance is assigned only one label. 
However, this single-label approach is not always practical. 
Attackers or threat groups behind malicious applications typically use a combination of different Tactics and Techniques to carry out their attacks. 
Thus, instead of the traditional single-label approach, we need to adopt a multi-label classification approach, where multiple labels are assigned to a single instance.
 Unlike a single-label dataset, which has only one column for labels, a multi-label classification dataset uses multiple columns for labels, with each column representing a different label. If two labels are associated with a single instance, the corresponding columns for those labels will have a value of 1 for that instance; otherwise, the value will be 0. 
To perform multi-label classification, we utilized the Problem Transformation Approach (PTA) \cite{tsoumakas2010mining}. This technique involves converting the multi-label classification task into several single-label classification problems. 
In our study, we used three PTAs that are commonly used in the literature. 
\begin{itemize}
    \item \textit{Binary Relevance:} Binary relevance decomposes the multi-label classification task into multiple independent single-label binary classification problems. Specifically, 
    for each label \( l_i \in \mathcal{L} \), a binary classifier \( h_i \) is trained to predict whether \( l_i \) is a part of \( Y_i \). The binary classifier \( h_i \) outputs 1 if the label \( l_i \) is present and 0 otherwise. Mathematically, each classifier learns a function \( h_i: X \rightarrow \{0, 1\} \). The overall multi-label prediction for a new instance \( x \) is obtained by combining the predictions of all binary classifiers:
\[ \hat{Y} = \{ l_i \in \mathcal{L} \mid h_i(x) = 1\} \]

\item  \textit{Classifier chains:} In this approach, instead of training independent binary classifiers, a sequence (or chain) of binary classifiers is trained. Each classifier in the chain uses not only the original features, but also the predictions of previous classifiers as additional features.  Each binary classifier \( h_i \) for label \( l_i \) predicts \( y_i \) using the feature vector \( x \) and the predictions of previous classifiers \( y_1, y_2, \ldots, y_{i-1} \). For example, \( x \in X \), the process starts with the original features \( x \). The first classifier \( h_1 \) predicts \( y_1 \) using \( x \). The second classifier \( h_2 \) predicts \( y_2 \) using \( x \) and \( y_1 \). This process continues until the last classifier \( h_t \) predicts \( y_t \) using \( x \) and all previous predictions. The final multi-label prediction for \( x \) is the set of all labels \( l_i \) for which \( y_i = 1 \). 

\item \textit{Label Powerset:} Instead of handling each label independently, Label Powerset (LP) considers every possible combination of labels as a single class, resulting in \( 2^t \) possible classes. For example, if there are three labels, \( \{l_1, l_2, l_3\} \), LP creates classes for every possible combination of these labels, such as \( \{l_1\} \), \( \{l_1, l_3\} \), \( \{l_2, l_3\} \), \( \{l_1, l_2\} \), etc. The classifier is then trained to predict one of these combinations for each instance. During prediction, the model outputs one of these pre-defined label combinations, which represents the set of labels assigned to the instance. 

\end{itemize}
For the implementation of these PTAs, we employ a diverse set of classifiers that are effective in multi-label classification. Random Forest (RF) \cite{wu2019multi} and Decision Trees (DT) \cite{vens2008decision} excel at modeling complex decision boundaries while leveraging ensemble methods to capture label correlations. XGBoost enhances performance in structured data by efficiently handling imbalanced multi-label distributions and optimizing multiple objectives through boosting techniques \cite{bohlender2020extreme}. Meanwhile, Multi-Layer Perceptron (MLP) leverages deep learning to learn hierarchical representations and capture label dependencies through hidden layers, improving predictive accuracy in multi-label scenarios \cite{cerri2014hierarchical}. We selected these classifiers due to their demonstrated effectiveness in previously published research on similar tasks.~\cite{sharma2023radar, kim2022comparative}.

\section{LLM Approaches}
In this section, we explore alternative LLM-based solutions to execute the second phase. Specifically, we introduce a Retrieval-Augmented solution and a Fine-tuning methodology.

\subsection{Prompt Engineering}
In this phase, we develop a prompt-based approach to predict TTPs from static features of Android applications using LLMs. Designing prompts that effectively guide LLMs in generating accurate and relevant responses is a challenging task \cite{chang2024survey}. A well-designed prompt should strategically direct the model to leverage its internal knowledge effectively. Prompt engineering involves designing inputs that align the model outputs with desired objectives \cite{chen2023unleashing}. The effectiveness of a prompt depends on multiple factors, including its structure, formatting, and linguistic nuances, as LLMs exhibit high sensitivity to subtle variations in these elements~\cite{yang2024harnessing}.
We employ prompt engineering to optimize the extraction of relevant TTPs. 
We implement four distinct prompting strategies, each tailored to extract TTPs from static features with varying degrees of user-defined instruction specificity. After evaluating the performance of these prompts, we integrate a RAG approach to further enhance the model accuracy and contextual relevance. In the following, we list the four prompt templates used in this study.

\begin{table}[ht]
    \centering
    \footnotesize
    \fcolorbox{black}{gray!15}{\parbox{\linewidth}{ 
    \fontfamily{pcr}\selectfont
\textbf{\textit{Instruction}:} Analyze the static features associated with an Android application. Identify and list the relevant MITRE Tactics and Techniques using the context provided.
\\

\textit{\textbf{Response Format:}}
\\
Tactic(s): \textless List of Tactics \textgreater

Technique(s): \textless List of Techniques \textgreater
\\

If the provided data is insufficient to determine the Tactics and Techniques, respond with:
"Not enough information."
\\

Question:\textbf{ \{\textit{question}\}}
}}
    \caption{Prompt 1: Basic prompt}
    \label{prompt1}
\end{table}

\begin{enumerate}
    \item  The first prompt strategy provides straightforward instruction to analyze the static features and predict the associated Tactics and Techniques, as shown in Table \ref{prompt1}. This prompt includes clear fundamental parameters for the task, a structured format for the output, and a fallback response, ``Not enough information'' if the data are insufficient for making predictions.

\begin{table}[ht]
    \centering
    \footnotesize
    \fcolorbox{black}{gray!15}{\parbox{\linewidth}{ 
    \fontfamily{pcr}\selectfont
\textbf{\textit{Instruction}:} You are a cybersecurity expert specializing in mobile application security. Your role is to analyze the static features associated with an Android application. Identify and list the relevant MITRE Tactics and Techniques using the context provided. Refer to this link for more details (\url{https://attack.mitre.org/matrices/mobile/android/})
\\

\textit{\textbf{Response Format:}}
\\
Tactic(s): \textless List of Tactics, e.g., Collection, Impact \textgreater

Technique(s): \textless List of Techniques, e.g., T1636, T1582, T1604, T1437, T1521, T1417 \textgreater
\\

If the provided data is insufficient to determine the Tactics and Techniques, respond with:
``Not enough information.''
\\

\textit{\textbf{Note:}} Only provide the Tactics and Techniques in the specified format. Do not include any additional explanations or comments.
\\

Question:\textbf{ \{\textit{question}\}}
}}
    \caption{Prompt 2: Expert level prompt}
    \label{prompt2}
\end{table}

\item The second prompting strategy (see Table \ref{prompt2}) introduces a domain-expert context by framing the query from the perspective of a cybersecurity expert. This helps to activate the domain-relevant knowledge of the model, potentially improving prediction accuracy through expert-aligned reasoning patterns. The prompt incorporates references to the MITRE ATT\&CK framework, includes a precise and professional output format, and uses domain-specific language to simulate expert-level analysis.
\begin{table}[h!]
    \centering
    \footnotesize
    \fcolorbox{black}{gray!15}{\parbox{\linewidth}{ 
    \fontfamily{pcr}\selectfont

\textbf{\textit{Instruction:}} Analyze the given static features associated with an Android application. Identify and list the relevant MITRE Tactics and Techniques linked to these features.
\\

Static Features: \{\textit{\textbf{question}}\}
\\

\textbf{Tactics:}  
Tactics represent the broad objectives that an adversary seeks to achieve. Identify the Tactics (one or many) from the MITRE ATT\&CK for Android matrix (\url{https://attack.mitre.org/matrices/mobile/android/}) associated with the static features.
\\

\textbf{Techniques:}  
Techniques are specific methods adversaries employ to achieve their Tactics. List the relevant technique IDs associated with the given static features. Some examples of Techniques are: T1636, T1582, T1604, T1437, etc. Refer MITRE ATT\&CK Android matrix site for technique IDs.
\\

\textbf{Scoring:}  
For each identified Tactic and Technique, assign a score from 0 to 5 based on the relevance and likelihood of the Tactic/Technique being linked to the given static features.
\\

\textbf{Response Format:}  \\
\textbf{Tactics and Scores:}   \\
- $<$Tactic 1$>$: $<$Score (0-5)$>$   \\
- $<$Tactic 2$>$: $<$Score (0-5)$>$   \\
... \\

\textbf{Techniques and Scores:}  \\
- $<$Technique 1$>$: $<$Score (0-5)$>$   \\
- $<$Technique 2$>$: $<$Score (0-5)$>$   \\
... \\
\\

Using the details provided, return the associated Tactics and Techniques with scores.
\\

If there is insufficient information to determine the Tactics and Techniques, respond with:  
\textit{"Not enough information."}
\\

\textbf{Note:}  
- Provide only the Tactics, Techniques, and scores in the specified format.  
- Do not include any additional explanations or comments.
\\
}}
    \caption{Prompt 3: Expert level prompt with score}
    \label{prompt3}
\end{table}

\item The third prompt represents an advanced strategy for analyzing Android application security by leveraging static features to identify associated MITRE ATT\&CK Tactics and Techniques as shown in Table \ref{prompt3}. It incorporates {\em(i)} a structured definition of Tactics and Techniques, {\em(ii)} explicit references to the MITRE ATT\&CK Android matrix for threat intelligence validation, {\em(iii)} a scoring mechanism to quantify the likelihood and relevance of identified TTPs, and {\em(iv)} a standardized response format to ensure consistency and interpretability in security assessments.
\begin{table}[h!]
\footnotesize
\centering
\resizebox{\textwidth}{!}{ 
\renewcommand{\arraystretch}{1.5} 
\setlength{\arrayrulewidth}{1pt} 
\setlength{\tabcolsep}{10pt} 
\fontfamily{pcr}\selectfont
\begin{tabular}{|>{\columncolor[gray]{0.9}}p{15.5cm}|}
\hline
\textbf{\textit{Instruction}}: You are a cybersecurity expert specializing in mobile application security. Your task is to respond to queries associated with Android applications. Each Android app has different static features.
\newline

\textit{\textbf{Understanding Static Features}}
\newline
Static features are characteristics of an Android application that can be analyzed without executing the app. These features provide insights into the app's behavior, permissions, and components.
\newline
Here are examples of static features commonly used in analyzing Android applications:
\newline
\newline
\textit{\textbf{Permissions}:} These are declarations that specify what resources the app can access. , e.g., android.permission.INTERNET, android.permission.ACCESS\_FINE\_LOCATION.
\newline
\textit{\textbf{Activities}:} Activities represent a single screen in an app's user interface., e.g., com.example.app.MainActivity.
\newline
\textit{\textbf{Services}:} Services running in the background, e.g., com.example.app.BackgroundSyncService.
\newline
\textit{\textbf{Intents}:} Specify how the app communicates with other apps or handles specific actions, e.g., android.intent.action.VIEW.
\newline
\textit{\textbf{Broadcast Receivers}:} Components listening for system or app-level broadcasts, e.g., com.example.app.BatteryLowReceiver.
\newline
\newline
By analyzing the static features, it is possible to infer the Tactics and Techniques used by adversaries to compromise mobile security. For more details on static features, you can refer to the [Android Developers documentation](\url{https://developer.android.com/guide/topics/manifest/manifest-intro}).
\newline
\newline
To conduct attacks through Android applications, adversaries use different Tactics and Techniques.
\newline
\newline
\textit{\textbf{Understanding Tactics}}
\newline
In the context of cybersecurity, 
Tactics refer to the general goals or objectives that an adversary aims to achieve through their actions. These can include a wide range of malicious activities, such as Collection, Impact, Defense Evasion, etc.
\newline
\newline
\textit{\textbf{Understanding Techniques}}
\newline
Techniques are specific methods or actions that adversaries use to achieve their Tactics. Each Tactic in the MITRE ATT\&CK framework is mapped to multiple Techniques. Some examples of Techniques are: T1636, T1582, T1604, T1437, etc.
\newline
\newline
More details about Tactics and Techniques are available at \textit{\textbf{MITRE ATT\&CK}} (\url{https://attack.mitre.org/matrices/mobile/android/}), which provides a matrix that maps Tactics to specific Techniques used by adversaries. Other threat intelligence platforms such as \textit{\textbf{VirusTotal}} (\url{https://www.virustotal.com/gui/home/upload}), and \textbf{Alienvault} (\url{https://otx.alienvault.com/}) also provide valuable insights into potential malicious behaviors and TTPs associated with various applications. Also, refer to TRAM (\url{https://github.com/mitre-attack/tram}) for more TTP mapping.
\newline
\newline
Using the details provided, analyze the static features of an Android application and identify the associated Tactics and Techniques. One app may be associated with multiple Tactics and Techniques.
\newline
\newline
\textit{\textbf{Response Format:}}
\newline
Tactic(s): \textless List of Tactics, e.g., Collection, Impact \textgreater
\newline
Technique(s): \textless List of Techniques, e.g., T1636, T1582, T1604, T1437, T1521, T1417 \textgreater
\newline
\newline
If the provided data is insufficient to determine the Tactics and Techniques, respond with: "Not enough information."
\newline
\newline
\textit{Note:} Only provide the Tactics and Techniques in the specified format. Do not include any additional explanations or comments.
\newline
\newline
Question: {\textbf{\textit{question}}} 
\\
\hline
\end{tabular}
}
\caption{Prompt 4: Detailed prompt}
\label{prompt4}
\end{table}

\item The fourth strategy represents the most advanced approach, incorporating comprehensive technical context about Android static features and their security implications, as shown in Table \ref{prompt4}. This prompt includes {\em(i)} detailed technical specifications of Android static features, {\em(ii)} definitions of Tactics and Techniques, {\em(iii)} references to threat intelligence platforms, and {\em(iv)} a standardized output schema optimized for consistency in predictions.
\end{enumerate}
\subsection{Retrieval Augmented Generation for Tactics and Techniques Prediction}
This section explores the effectiveness of the RAG model in predicting Tactics and Techniques based on static features. As explained in the background \ref{sec:background}, the RAG framework comprises two primary components: the \textit{retriever} and the \textit{generator}. The retriever is tasked with locating relevant information within a large knowledge base, which is stored in a vector database. This database holds precomputed dense vector representations of the indexed information. To generate the knowledge base, we utilized the same training samples as in the ML-based approach but stored them differently in CSV format. Each training sample comprises three key elements: {\em(i)} Description represents a descriptive summary of the app's features, formatted as follows:
 \textit{``Activities related to app \textless Hash \textgreater are: \textless Activities \textgreater. Permissions required: \textless Permissions \textgreater. Services used: \textless Services \textgreater. Receivers included: \textless Receivers \textgreater. Intent Actions: \textless Intent Actions \textgreater and Intent Categories: \textless Intent Categories \textgreater''}. {\em(ii)} Tactic represents the associated Tactics used by the attacker using this particular app {\em(iii)} Technique specifies the relevant Techniques associated with the attack.
 
We use the CSVLoader to load the dataset and divide it into smaller, manageable segments for easier processing. Each segment is then transformed into a dense vector representation using an embedding model. Once converted into dense vectors, the dataset is indexed in a vector database. This indexing facilitates efficient similarity searches, enabling the retrieval of the most contextually relevant documents in response to queries.

When a mobile security analyst submits a query, it undergoes processing and conversion into a vector representation using the same embedding model. This vector representation is then used with an Approximate Nearest Neighbor algorithm (ANN) to search for the most similar documents in the vector database, using cosine similarity as the metric. The ANN algorithm ensures that only the top-$k$ most relevant documents are retrieved, reducing the computational burden while ensuring that the most pertinent context is available for generating predictions. Once the retriever identifies the top-$k$ relevant documents, these, along with the query, are passed to the generator component. The generator, typically a pretrained LLM, is responsible for producing the final output based on the provided context and prompt. Specifically, the LLM is given the relevant context retrieved by the retriever, reducing the likelihood of hallucinations (i.e., generating incorrect or fabricated information). In our approach, the prompt used for the generator is the same prompt identified in the previous phase as the most effective for generating accurate responses. This methodology significantly mitigates the risk of irrelevant or fabricated information, as the model relies solely on the context retrieved from the vector database.

\subsection{Fine Tuning LLM}
When adapting pre-trained LLMs for specific applications, fine-tuning avoids the resource-intensive process of complete parameter retraining. In the context of cybersecurity, a significant challenge arises from the lack of domain-specific information in pre-training datasets, leading to hallucination issues. Fine-tuning addresses this gap by incorporating specialized knowledge, thereby reducing hallucinations and improving accuracy in predicting attackers' Tactics and Techniques.
In this study, we experimented with a range of open-source pre-trained LLMs with various strengths and architectures, including SecBERT \footnote{\url{https://github.com/jackaduma/SecBERT}} and CySecBERT \cite{bayer2024cysecbert}, which were developed specifically for cybersecurity tasks. We selected \textit{Phi-3-mini-4k-instruct} for its ability to generalize well, despite its compact size. The Phi series of Small Language Models (SLMs) is notable for its unique combination of high performance and cost-effectiveness, consistently outperforming models of similar or larger size in various linguistic tasks \cite{abdin2024phi}. Moreover, we experimented with \textit{Meta-Llama-3-8B-Instruct} for its versatile capabilities and efficient architecture, which are used in various downstream tasks \cite{kumar2024prompt}. This $8$-billion parameter model strikes an optimal balance between computational requirements and performance capabilities. Similarly, we experimented with \textit{Mistral-7B-Instruct-v0.2} for its optimal performance and efficiency \cite{fieblinger2024actionable}.

To optimize the fine-tuning process, we employed Quantized Low-Rank Adaptation (QLoRA) \cite{dettmers2024qlora} Technique with $4$-bit quantization. This approach minimizes computational demands while maintaining model effectiveness, making it particularly valuable for updating large models with limited computing resources.

\section{Experiment and Evaluation}
\label{sec:result}
This section describes the experiments carried out to test the effectiveness and robustness of our approach.
In particular, in the following sections, we will describe the testbeds, the data, and the evaluation metrics used to test our framework, and we report the obtained results along with their critical analysis.

\subsection{Experimental Setup}
We conducted the implementation on an Intel i9 Windows system equipped with a 5GB GPU. We also used a workstation equipped with an Intel i9-14900KF, dual RTX4090 GPUs, and 128 GB of RAM for LLM experimentation. For the fine-tuning process, we leveraged the Hugging Face Transformers library, utilizing pre-trained language models and optimizing them with PyTorch. In the RAG framework, we integrated Facebook AI Similarity Search (FAISS)\footnote{\url{https://ai.meta.com/tools/faiss}} for efficient document retrieval and LangChain \footnote{\url{https://www.langchain.com}} to manage the retrieval and response generation pipeline. For multi-label classification tasks, we employed the scikit-multilearn library. To mitigate class imbalance, we incorporated MLSMOTE \cite{charte2015mlsmote} for the generation of synthetic samples. To enhance model interpretability, we utilized the SHAP library and, for data visualization, we used the Python library matplotlib. 
\begin{figure}
    \centering
    \includegraphics[scale=0.31]{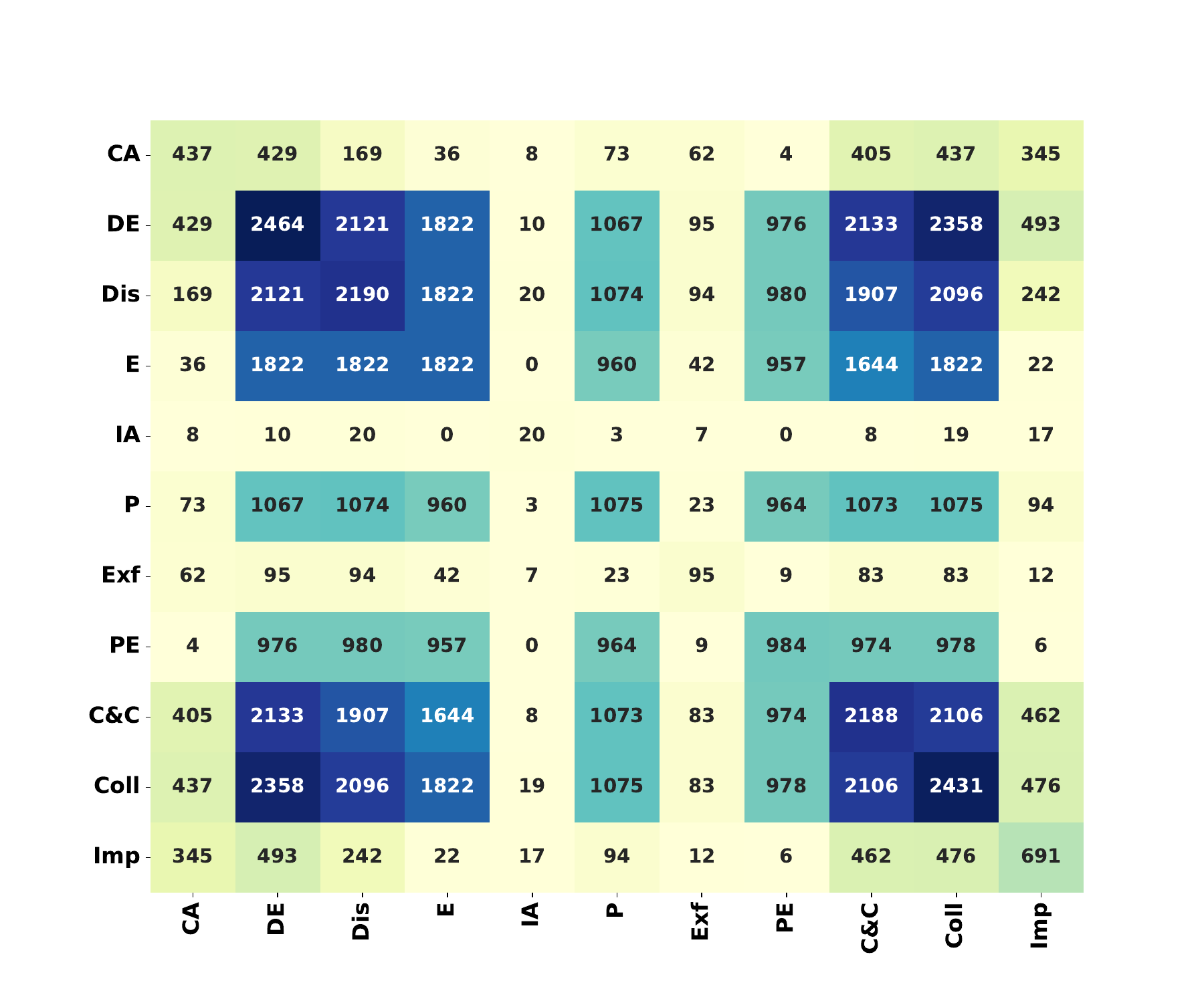}
    \caption{Frequency of samples across various classes in the Tactic dataset. \textbf{CA-} Credential Access, \textbf{DE-} Defense Evasion, \textbf{Dis-} Discovery,\textbf{ E-} Execution, \textbf{IA-} Initial Access, \textbf{P-} Persistence, \textbf{Exf-} Exfiltration, \textbf{PE-} Privilege Escalation, \textbf{C\&C-} Command and Control, \textbf{Coll-} Collection, \textbf{Imp-} Impact.}
    \label{fig:tactic_Statistics}
\end{figure}
\subsection{Dataset}
\label{subsec:dataset}
\begin{figure}
    \centering
    \includegraphics[width=0.6\linewidth]{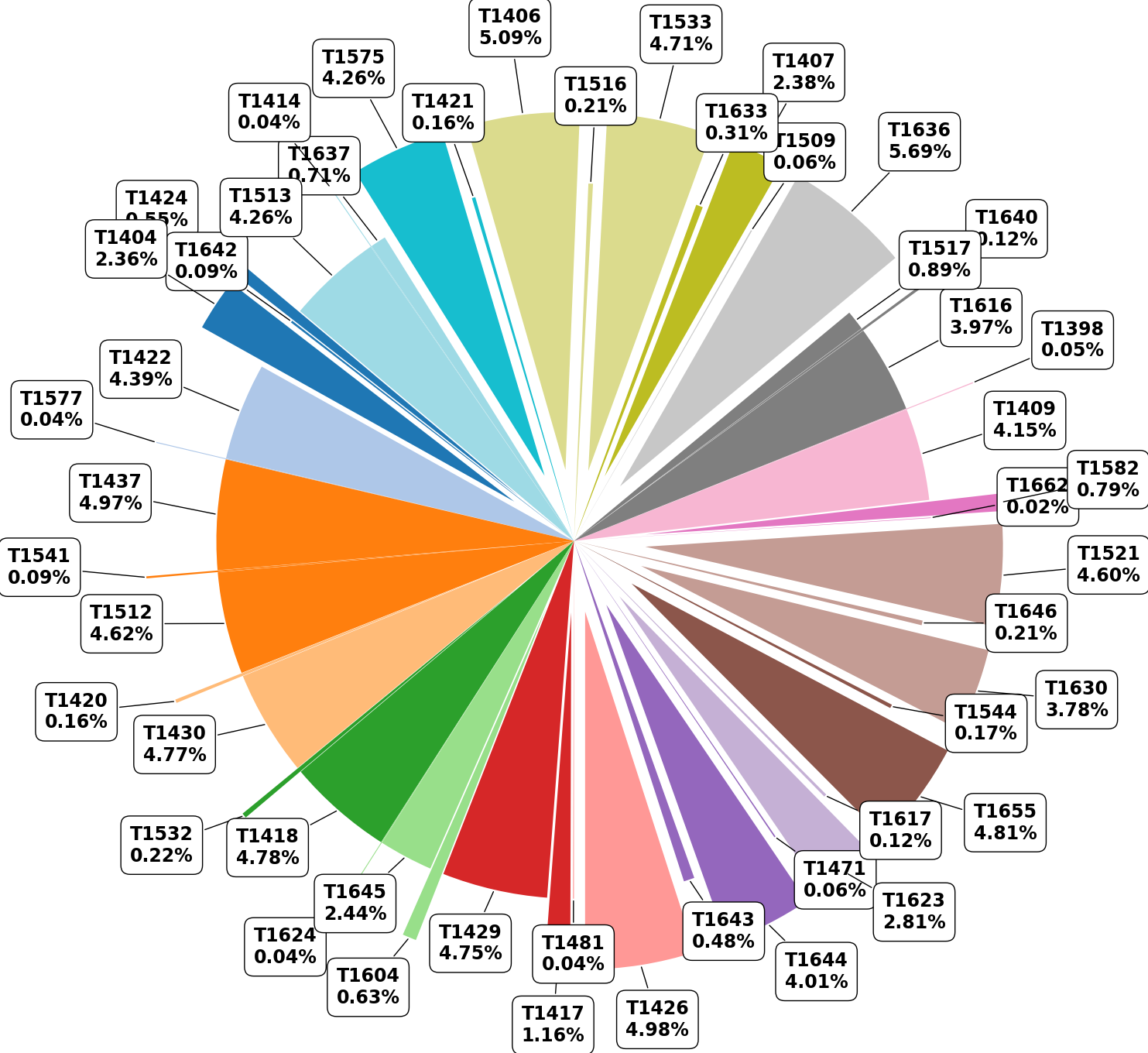}
    \caption{Technique ID Distribution in DroidTTP Technique Dataset}
    \label{fig:technique_data_statistics}
\end{figure}

As discussed in Section \ref{sub:collection}, to the best of our knowledge, no existing dataset was available. Therefore, we created our own dataset for experimentation following the process described in Section \ref{sec:methodology}. We initially extracted $3034$ hashes from the procedure references from the MITRE ATT\&CK knowledge base. These hashes were subsequently verified using VirusTotal, which revealed that $261$ hashes did not correspond to Android applications. Consequently, these hashes were excluded from further experimentation. We excluded the \texttt{Lateral Movement} Tactic label from the dataset due to its association with only a single hash, which lacks sufficient data for effective model training and generalization. After completing all necessary steps, we compile a Tactic classification dataset consisting of $2774$ apps with $13324$ features and $11$ Tactic labels. The distribution of the samples from the Tactic data set is illustrated in Figure \ref{fig:tactic_Statistics}. Since the dataset is multi-labeled, an app can exhibit multiple Tactics. \ref{fig:tactic_Statistics} report the number of samples for each Tactic class as well as the co-occurrence of samples across multiple Tactic classes. For example, the dataset includes $2,464$ samples labeled with the \texttt{Defense Evasion} Tactic and $20$ samples labeled with the \texttt{Initial Access} Tactic. This indicates that the dataset is highly imbalanced. Furthermore, the figure reveals that the $429$ samples are labeled with both the \texttt{Credential Access} and \texttt{Defense Evasion} Tactics. Similarly, we generated a Technique classification dataset with the same number of samples and feature sets. The distribution of techniques is illustrated in Figure~\ref{fig:technique_data_statistics}. Specifically, most techniques are associated with a similar percentage of samples, with T1636 and T1406 appearing most frequently. In contrast, techniques such as T1481 and T1662 occur at significantly lower rates.  
While MITRE ATT\&CK defines $85$ Techniques for Android, many of these techniques lack sufficient malware hash data for effective model training. Consequently, we were able to collect adequate data for only $48$ Techniques. As a result, our Technique classification task focuses on these $48$ Techniques, ensuring optimal model performance, accuracy, and generalizability within the available data scope. 
\begin{table}[h!]
\caption{Optimized Hyperparameters for Tactic and Technique Classification}
    \centering
    \footnotesize
    \resizebox{\textwidth}{!}{
    \begin{tabular}{clp{7cm}p{7cm}}
    \toprule
        \textbf{PTA} & \textbf{Classifier} & \textbf{Tactic} & \textbf{Technique} \\
        \midrule
        \multirow{4}{*}{BR} & DT & criterion: entropy, max\_depth: 41, min\_samples\_leaf: 3, min\_samples\_split: 17 & criterion: gini, max\_depth: 63, max\_leaf\_nodes: 50, min\_samples\_leaf: 3, min\_samples\_split: 12\\
        \cline{2-4}
         & RF & max\_depth: None, min\_samples\_leaf: 1, min\_samples\_split: 3, n\_estimators: 78 & max\_depth: 20, min\_samples\_leaf: 2, min\_samples\_split: 3, n\_estimators: 64\\
        \cline{2-4}
         & XGBoost & gamma: 0.632, learning\_rate: 0.242, max\_depth: 9, n\_estimators: 82 &  gamma: 1.679, learning\_rate: 0.124, max\_depth: 14, n\_estimators: 156\\
        \cline{2-4}
         &MLP & hidden\_layer\_sizes: 100, activation: relu, solver: Adam, learning\_rate: 0.001 & hidden\_layer\_sizes: 100, activation: relu, solver: Adam, learning\_rate: 0.001 \\
        \midrule
        \multirow{4}{*}{CC} & DT & criterion: entropy, max\_depth: 91, max\_leaf\_nodes: 100, min\_samples\_leaf: 2, min\_samples\_split: 14 & criterion: gini, max\_depth: 65, max\_leaf\_nodes: 20, min\_samples\_leaf: 2, min\_samples\_split: 5 \\
        \cline{2-4}
         & RF &  max\_features: auto, min\_samples\_leaf: 1, min\_samples\_split: 5, n\_estimators: 96& max\_features: auto, min\_samples\_leaf: 1, min\_samples\_split: 5, n\_estimators: 178\\
        \cline{2-4}
         & XGBoost &  gamma: 0.644, learning\_rate: 0.243, max\_depth: 5, n\_estimators: 191&  gamma: 1.896, learning\_rate: 0.253, max\_depth: 12, n\_estimators: 54 \\
        \cline{2-4}
         &MLP & hidden\_layer\_sizes: 100, activation: relu, solver: Adam, learning\_rate: 0.001&  hidden\_layer\_sizes: 100, activation: relu, solver: Adam, learning\_rate: 0.001\\
        \hline
        \multirow{4}{*}{LP} & DT & criterion: gini, max\_depth: 31, max\_leaf\_nodes: 50, min\_samples\_leaf: 4, min\_samples\_split: 19 & criterion: entropy, max\_depth: 46, max\_leaf\_nodes: None, min\_samples\_leaf: 3, min\_samples\_split: 4\\
        \cline{2-4}
         & RF & max\_features: log2, min\_samples\_leaf: 1, min\_samples\_split: 6, n\_estimators: 199 & max\_features: auto, min\_samples\_leaf: 1, min\_samples\_split: 5, n\_estimators: 153 \\
        \cline{2-4}
         &XGBoost &  gamma: 0, learning\_rate: 0.3, max\_depth: 6, n\_estimators: 100& gamma: 0, learning\_rate: 0.3, max\_depth: 6, n\_estimators: 100\\
        \cline{2-4}
         &MLP & hidden\_layer\_sizes: 100, activation: relu, solver: Adam, learning\_rate: 0.001& hidden\_layer\_sizes: 100, activation: relu, solver: Adam, learning\_rate: 0.001 \\
        \bottomrule
    \end{tabular}}
    \label{tab:hyper_parameter}
\end{table}

\subsection{Hyperparameter Setup}
The performance of the model is highly dependent on the hyperparameters selected for each classifier. To optimize these parameters, we applied Randomized Search Cross-Validation (CV) to the Decision Tree, Random Forest, and XGBoost classifiers. The optimized hyperparameters for each classifier, tailored for both Tactic and Technique classification tasks, are summarized in Table \ref{tab:hyper_parameter}.

\subsection{Evaluation Metrics}
To assess the performance of our Tactic and Technique classification model, we employed the evaluation metrics, such as Accuracy (A),  Weighted Precision (P), Weighted Recall (R), Weighted F1 score (F1), Jaccard Similarity (JS), and Hamming Loss (HL). 
\begin{itemize}
    \item \textit{Accuracy (A).} It measures the proportion of samples that were correctly classified by the model. 
    \begin{equation}
\text{A} = \frac{1}{N} \sum_{i=1}^{N} (y_i = \hat{y}_i)
    \label{eq:accuracy}
    \end{equation}
    where N is the total number of instances, $y_i$ is the true label for the $i$-th instance, and $\hat{y}_i$ is the predicted label for the $i$-th instance.

    \item \textit{Weighted Precision (P).} Precision evaluates the proportion of correctly predicted labels among all predicted labels. Weighted precision adjusts for class imbalance by computing the metric for each label and weighting it by the number of true instances of that label.
    \begin{equation}
        \text{P} = \frac{\sum_{t=1}^{T} w_t \cdot \text{Precision}_t}{\sum_{t=1}^{T} w_t}
    \end{equation}
    where, $t$ represents the index of each individual label, $T$ denotes the total number of unique labels, $\text{Precision}_t = \frac{\text{TP}_t}{\text{TP}_t + \text{FP}_t} $ and $w_t = \text{TP}_t + \text{FN}_t$

    \item \textit{Weighted Recall (R).} It measures the proportion of correctly predicted labels among all true labels.
    \begin{equation}
        \text{R} = \frac{\sum_{t=1}^{T} w_t \cdot \text{Recall}_t}{\sum_{t=1}^{T} w_t}
    \end{equation}
    where, $\text{Recall}_t = \frac{\text{TP}_t}{\text{TP}_t + \text{FN}_t} $

    \item \textit{Weighted F1-score (F1).} It is the harmonic mean of precision and recall. 
    \begin{equation}
        \text{F1} = \frac{\sum_{t=1}^{T} w_t \cdot \text{F1}_t}{\sum_{t=1}^{T} w_t}
    \end{equation}
    where, $\text{F1}_t = 2 \cdot \frac{\text{P}_t \cdot \text{R}_t}{\text{P}_t + \text{R}_t} $

    \item \textit{Jaccard Similarity (JS).} It measures how similar the predicted set of labels is to the true set of labels. A higher Jaccard Similarity indicates better model performance.
\begin{equation} 
\text{JS} = \frac{1}{n} \sum_{i=1}^{n} \frac{| y_i \cap \hat{y}_i |}{| y_i \cup \hat{y}_i |} 
\label{eq_js}
\end{equation}
where \( n \) is the total number of instances. \( | y_i \cap \hat{y}_i | \)  is the size of the intersection of the true and predicted label sets (i.e., the number of labels that are correctly predicted). \( | y_i \cup \hat{y}_i | \) is the size of the union of the true and predicted label sets (i.e., the total number of unique labels in either the true or predicted set).

\item \textit{Hamming Loss (HL).} Measures the average fraction of labels that are incorrectly predicted across all instances. \textit{Hamming Loss (HL)} is calculated using the Equation \ref{eq:hamming_loss}. A lower Hamming Loss corresponds to better performance, with \( HL = 0 \) indicating perfect predictions (i.e., all labels are correctly predicted) and \( HL = 1 \) meaning that all labels are incorrectly predicted for all instances.
\begin{equation}
   HL = \frac{1}{n} \sum_{i=1}^{n} \frac{1}{|\mathcal{L}_i|} \sum_{j=1}^{|L_i|} \mathbb{I}(y_{ij} \neq \hat{y}_{ij})
\label{eq:hamming_loss}
\end{equation}
where \( n \) is the total number of instances. \( |\mathcal{L}_i| \) is the number of labels for the \( i \)-th instance. \( y_{ij} \) is the actual label of the \( j \)-th label for the \( i \)-th instance. \( \hat{y}_{ij} \) is the predicted label of the \( j \)-th label for the \( i \)-th instance. \( \mathbb{I}(\text{condition}) \) is the indicator function, returning 1 if the condition is true and 0 otherwise.
\end{itemize}

\subsection{Experimental Evaluation of Tactic Identification}
In this section, we present the results of Tactic classification experiments.
As discussed in Section \ref{sec:related_work}, we experimented with various PTAs and ML algorithms. Initially, we developed the Tactic model using the full feature set and tested it with ten different random seeds. To evaluate the performance of the Tactic models, we calculated the average score across these random seeds. Table \ref{tab:tactc_result_full_feature} summarizes the performance of the Tactic models. The findings indicate that the Label Powerset method combined with XGBoost demonstrates the highest performance compared to other classifiers. This model achieves a low Hamming Loss of 0.0102, a high Jaccard Similarity index of 0.9794, and an F1-score of 0.9892.

\begin{table}[h!]
\caption{Performance comparison of Tactic classification models}
    \centering
    \scriptsize
    \begin{tabular}{clcccccc}
    \toprule
        \textbf{PTA} & \textbf{Classifier} & \textbf{A} & \textbf{P} & \textbf{R} & \textbf{F1} & \textbf{JS} & \textbf{HL} \\
\midrule
        \multirow{4}{*}{BR} & DT & 0.9218 & 0.9833 & 0.9829 & 0.9828 & 0.9630 & 0.0159 \\
    
         & RF & 0.9191 & 0.9831 & 0.9821 & 0.9817 & 0.9634 & 0.0164\\

         & XGBoost &  0.9539 & 0.9883 & 0.9906 & 0.9894 & 0.9765 & 0.0100\\

         &MLP &  0.9247 & 0.9833 & 0.9798 & 0.9811 & 0.9610 & 0.0173\\
        \midrule
        \multirow{4}{*}{CC} & DT & 0.9286 & 0.9795 & 0.9836 & 0.9812 & 0.9630 & 0.0175\\

         & RF & 0.9250 & 0.9819 & 0.9842 & 0.9822 & 0.9652 & 0.0160\\

         & XGBoost & 0.9618 & 0.9866 & 0.9914 & 0.9889 & 0.9779 & 0.0104\\

         &MLP & 0.9240 & 0.9847 & 0.9782 & 0.9810 & 0.9596 & 0.0173\\
        \midrule
        \multirow{4}{*}{LP} & DT & 0.9587 & 0.9840 & 0.9876 & 0.9856 & 0.9730 & 0.0135\\

         & RF & 0.9623 & 0.9851 & 0.9875 & 0.9861 & 0.9749 & 0.0130\\

         &XGBoost & \textbf{0.9694} & \textbf{0.9880} & \textbf{0.9905} & \textbf{0.9892} & \textbf{0.9794} & \textbf{0.0102}\\

         &MLP & 0.9310 & 0.9759 & 0.9728 & 0.9737 & 0.9523 & 0.0242\\
        \bottomrule
    \end{tabular}
    \label{tab:tactc_result_full_feature}
\end{table}

\par
\begin{figure}[!ht]
    \centering
    \includegraphics[width=1.1\linewidth]{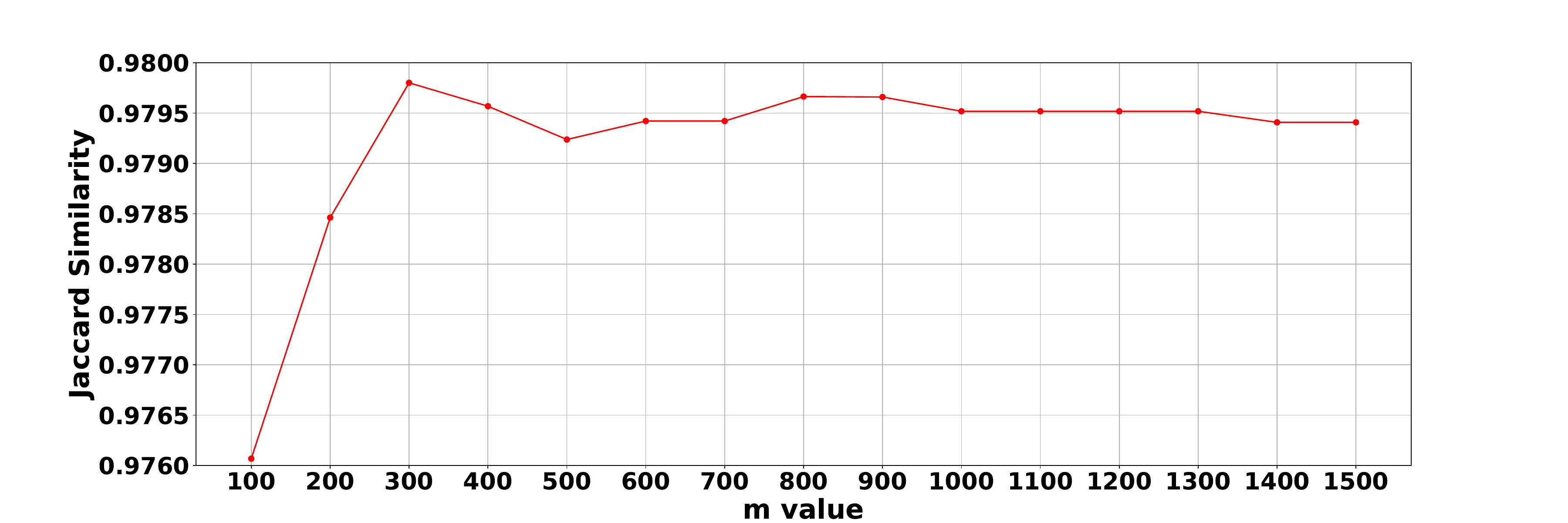}
    \caption{Jaccard Similarity score of Tactic classification model when feature selection applied.}
    \label{fig:tactic_js_feature}
\end{figure}

 We then applied the feature selection procedure outlined in Section \ref{sec:methodology} to reduce the feature set, optimizing computational efficiency and minimizing training time. As discussed, when the Tactic model was trained using the full feature set, the Label Powerset combined with XGBoost yielded the best performance; therefore, we applied feature selection specifically for the XGBoost classifier using the Label Powerset approach. As described in the feature selection procedure, we initially selected the top $m$ features for each class, then combined these top $m$ features across all labels. This process was repeated for $10$ different random seeds to identify the final top features across the various datasets generated by random seeds and labels. To achieve this, we tested with values of $m$ ranging from $100$ to $13,300$ and recorded the training time, top features, and various evaluation metrics. Figure \ref{fig:tactic_js_feature} presents the Jaccard Similarity score for the Tactic classification model after applying feature selection. 
We limited the plot to \( m \) values between 100 and 1500 after determining that beyond 1500 features, the results remained consistent when \( m \) was fixed at 1500.

From Figure \ref{fig:tactic_js_feature}, it is evident that the highest Jaccard Similarity score of $0.9798$ is achieved when $m$ is set to $300$. When the top $300$ features were selected for each label, we obtained $1737$ final features aggregated from all ten datasets. Using this optimized feature set, we developed the Tactic classification model, which is trained and evaluated across $10$ random seeds. This model is slightly higher than that of the model using the full feature set. Also, we observed that building a Tactic model using the full feature set required significantly longer execution time, taking $126.69$ seconds. In contrast, the optimized feature set of $300$ features reduced the execution time to just $18.61$ seconds. This substantial reduction in training time demonstrates the effectiveness of feature selection in significantly enhancing model efficiency without compromising performance.
\begin{table}[h!]
\caption{Performance comparison between original and augmented datasets for Tactics.}
\footnotesize
\centering
\begin{tabular}{lccc|ccc}
\toprule
\multirow{2}{*}{\textbf{Tactic}} & \multicolumn{3}{c|}{\textbf{Original Dataset}} & \multicolumn{3}{c}{\textbf{Augmented Dataset}} \\
\cline{2-7}
 & \textbf{P} & \textbf{R} & \textbf{F1} & \textbf{P} & \textbf{R} & \textbf{F1} \\
\midrule
\textbf{Credential Access}         & 0.95 & 0.97 & 0.96 & 0.99 & 0.99 & 0.99 \\
\textbf{Defence Evasion}           & 0.99 & 0.99 & 0.99 & 1.00 & 1.00 & 1.00 \\
\textbf{Discovery}                 & 0.99 & 0.99 & 0.99 & 0.99 & 1.00 & 1.00 \\
\textbf{Execution}                 & 0.99 & 1.00 & 1.00 & 0.99 & 1.00 & 1.00 \\
\rowcolor{lightgray} \textbf{Initial Access}            & 0.95 & 0.74 & 0.82 & 0.99 & 1.00 & 0.99 \\
\textbf{Persistance}               & 0.99 & 1.00 & 0.99 & 0.99 & 1.00 & 0.99 \\
\rowcolor{lightgray} \textbf{Exfiltration}              & 0.93 & 0.85 & 0.89 & 0.99 & 1.00 & 1.00 \\
\textbf{Privilage Escalation}      & 0.99 & 0.99 & 0.99 & 0.99 & 0.99 & 0.99 \\
\textbf{Command and Control}       & 0.99 & 0.99 & 0.99 & 0.99 & 1.00 & 0.99 \\
\textbf{Collection}                & 0.99 & 0.99 & 0.99 & 1.00 & 1.00 & 1.00 \\
\textbf{Impact}                    & 0.98 & 0.97 & 0.97 & 0.99 & 0.98 & 0.98 \\
\bottomrule
\end{tabular}

\label{tab:tactics_class_comparison}
\end{table}
To further analyze the classification results, we presented the individual weighted precision, recall, and F1 scores in Table \ref{tab:tactics_class_comparison}. From this table, we can infer that most Tactic classes achieved high performance, with scores exceeding $0.95$ across all metrics, except for the \texttt{Initial Access} and \texttt{Exfiltration} Tactics. The F1 score for \texttt{Initial Access} is $0.82$, while the F1 score for \texttt{Exfiltration} is $0.89$. This performance gap is primarily due to the low number of samples for these classes. Specifically, \texttt{Initial Access} has only $20$ samples, and the \texttt{Exfiltration} Tactic is represented by just $95$ samples, as illustrated in the Figure \ref{fig:tactic_Statistics}. So, to address this issue, we augmented the dataset using the  Multi-Label Synthetic Minority Over-sampling Technique (MLSMOTE) \cite{charte2015mlsmote}.

\begin{figure}[ht]
    \centering
    \includegraphics[width=0.9\linewidth]{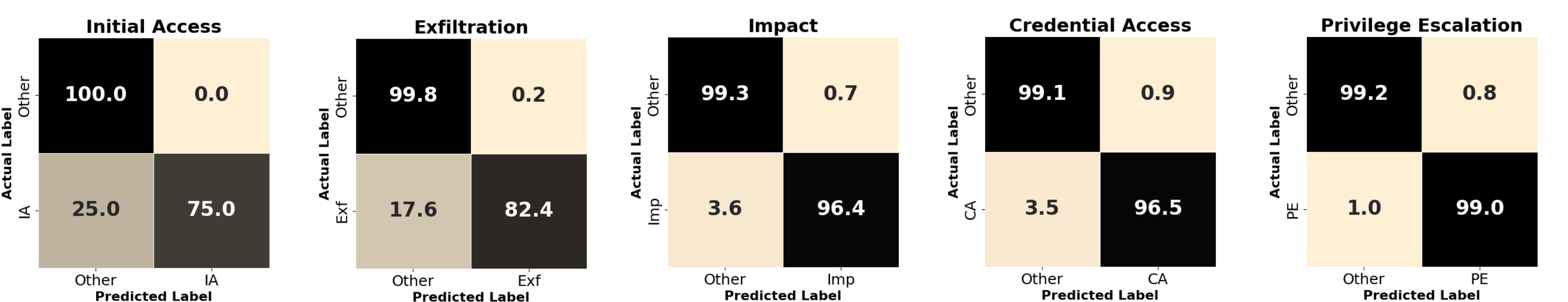}
    \caption{Confusion matrix of Tactic classification model before Augmentation}
    \label{fig:tactic_confusion_matrix_before_Augmentation}
\end{figure}

\begin{figure}[ht]
    \centering
    \includegraphics[width=0.9\linewidth]{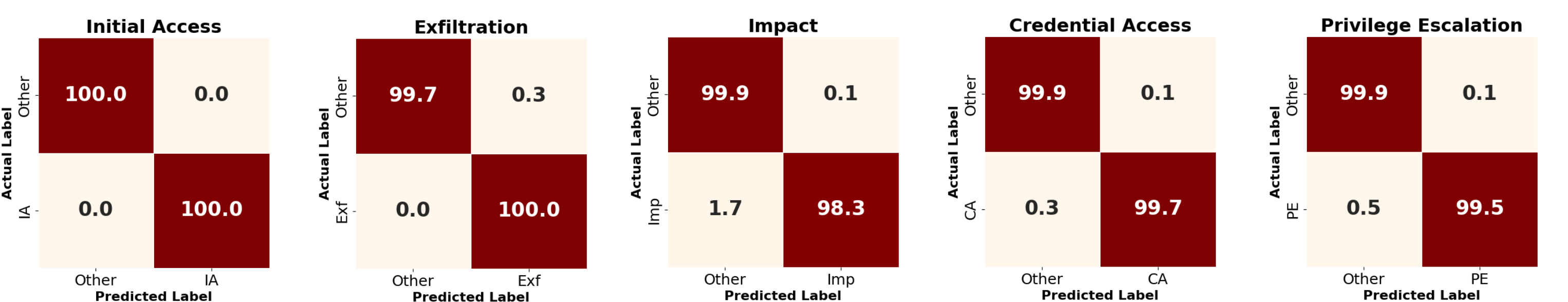}
    \caption{Confusion matrix of Tactic classification model after augmentation}
    \label{fig:tactic_confusion_matrix_after_augmentation}
\end{figure}
MLSMOTE is an adaptation of the popular Synthetic Minority Over-sampling Technique (SMOTE) \cite{chawla2002smote}, specifically developed for multi-label classification problems. In this technique, minority labels, such as \texttt{Initial Access} and \texttt{Exfiltration}, are identified based on their frequency within the dataset. After recognizing the minority labels, MLSMOTE identifies the minority instances, which are the data points associated with these underrepresented labels. To address the class imbalance, MLSMOTE generates synthetic samples for these minority labels. For each minority instance, it identifies the $k$-nearest neighbors within the same label group (we used the default value of $5$ nearest neighbors as specified in MLSMOTE). It then creates synthetic data by interpolating between the feature vectors of the minority instance and one of its neighbors, adding minor random variations. The synthetic instance is assigned the same labels as the original minority instance, and this process is repeated for other minority instances. To control the number of synthetic samples generated, we tested various sample sizes on the basis of the maximum count of the majority class. As shown in Figure \ref{fig:tactic_Statistics}, \texttt{Defense Evasion} has the highest count, with $2464$ instances. For this, we generated synthetic samples at rates of $25\%$, $50\%$, $75\%$, and $100\%$ of the highest class count, ensuring that the synthetic data did not exceed the original number of samples. 

\begin{table}[h!]
\footnotesize
\centering
\caption{Tactic model performance metrics at different percentages of MLSMOTE}
\begin{tabular}{ccccccc}
\toprule
\textbf{\%} & \textbf{A} & \textbf{P} & \textbf{R} & \textbf{F1} & \textbf{JS} & \textbf{HL} \\ 
\midrule
Original & 0.9694 & 0.9881 & 0.9908 & 0.9894 & 0.9798 & 0.0099 \\
25 & 0.9768 & 0.9912 & 0.9936 & 0.9924 & 0.9857 & 0.0074 \\ 
50 & 0.9784 & 0.9920 & 0.9941 & 0.9930 & 0.9868 & 0.0068 \\ 
75 & 0.9807 & 0.9933 & 0.9947 & 0.9940 & 0.9883 & 0.0058 \\ 
100 & \textbf{0.9819} & \textbf{0.9939} & \textbf{0.9950} & \textbf{0.9945} & \textbf{0.9893} & \textbf{0.0054} \\ 
\bottomrule
\end{tabular}
\label{tab:tactic_MLSMOTE_performance_metrics}
\end{table}
\begin{figure}[ht]
    \centering
    \subfigure[Exfilitration]{\includegraphics[width=0.48\linewidth]{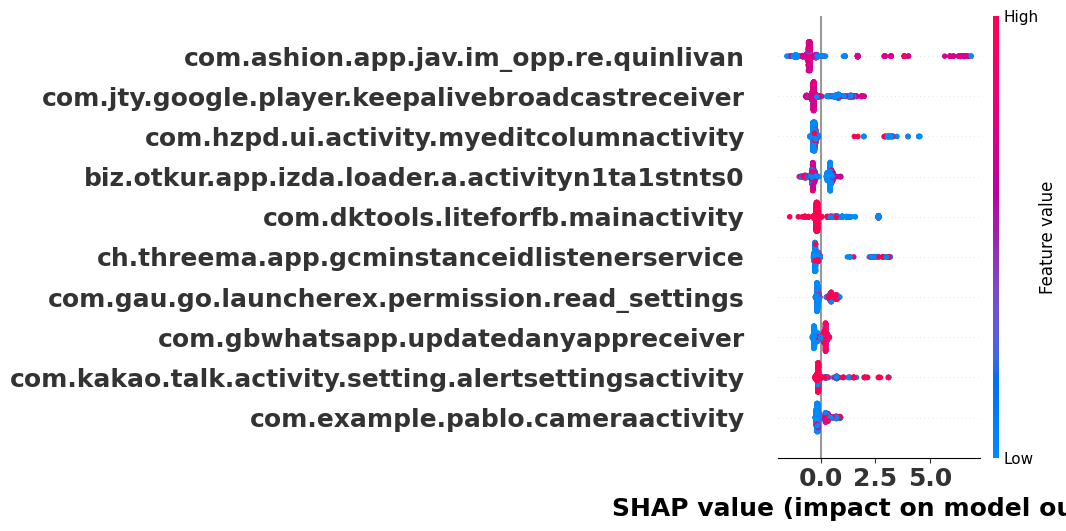}}
    \subfigure[Collection]{\includegraphics[width=0.48\linewidth]{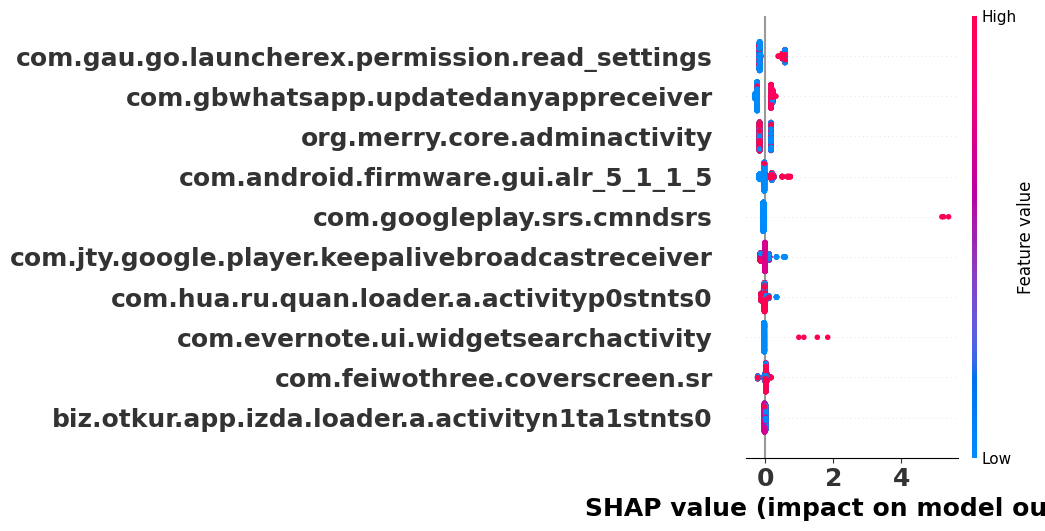}}
    \caption{SHAP summary plot of Tactic model}
    \label{fig:tactic_shap}
\end{figure}
After augmenting the dataset, we generated the Label Powerset using the XGBoost model with $10$ random seeds. The average results of the evaluation metrics for various sampling scenarios, along with those for the original dataset, are presented in Table \ref{tab:tactic_MLSMOTE_performance_metrics}. From the table, it is clear that as the number of synthetic samples increases, the model performance also improves. The best results were achieved when the number of synthetic samples matched the highest class count. In this case, the model achieved an accuracy of $0.9819$, precision of $0.9939$, recall of $0.9950$, F1-score of $0.9945$, Jaccard Similarity of $0.9893$, and a Hamming Loss of $0.0054$. Furthermore, our analysis shows that after data augmentation, the performance of all classes, particularly the minority classes such as \texttt{Initial Access} and \texttt{Exfiltration}, has significantly improved. As shown in Table \ref{tab:tactics_class_comparison}, the F1- score for \texttt{Initial Access} increased from $0.82$ to $0.99$, while the F1 score for \texttt{Exfiltration} rose from $0.89$ to $1.00$. These improvements highlight that the augmented dataset enhances the effectiveness of Tactic classification.

To further analyze the classification results, we plotted a confusion matrix, shown in Figure \ref{fig:tactic_confusion_matrix_before_Augmentation}. Since we used the Problem Transformation Approach, this led to multiple binary classification confusion matrices, one for each class label. As a result, the overall confusion matrix consists of $11$ individual matrices, each representing the classification performance for a specific Tactic. In the figure, we presented only $5$ confusion matrices corresponding to the lowest performance in Tactic classification using the original data. Additionally, we plotted the confusion matrices for the same classes after augmentation (Figure \ref{fig:tactic_confusion_matrix_after_augmentation}) to assess the impact on classification and misclassification rates. From these two confusion matrices, it is evident that the misclassification rate decreased after data augmentation. Specifically, before augmentation, the misclassification rate for \texttt{Initial Access} was $25\%$, which dropped to 0 after augmentation. Similarly, for the \texttt{Exfiltration} class, the misclassification rate decreased from $17.6\%$ to $0$. This pattern of improvement is observed across the other classes as well.

To thoroughly analyze and interpret the decision-making process of the model, we utilized the SHapley Additive exPlanations (SHAP) tool. SHAP provides granular insights into the contribution of each feature to the model's predictions, offering a comprehensive understanding of how static features influence the outcomes. We generated SHAP summary plots for each Tactic, with specific illustrations for the \texttt{Exfiltration} and \texttt{Collection} Tactics shown in Figure \ref{fig:tactic_shap}. In these plots, each dot corresponds to a SHAP value representing an individual dataset sample. Features are ranked according to their average impact on the model predictions, as indicated on the vertical axis. The horizontal axis denotes the SHAP values, reflecting the extent to which a feature drives the prediction towards or away from the target class. The color gradient ranging from blue to red reflects the actual feature value for each instance, with blue indicating a lower value and red indicating a higher value.

 In the SHAP summary plot for the \textit{Exfiltration} Tactic features such as \texttt{com.ashion.app.jav. im\_opp.re.quinlivan} and \texttt{com.jty.google.player.keepalivebroadcastreceiver} demonstrate a strong association with the model's prediction of exfiltration behavior. Higher SHAP values for these features (indicated by red dots) suggest that these specific app components are influential in predicting data exfiltration. For example, the feature \texttt{com.jty.google.player.keep alivebroadcastreceiver} represents a background service that ensures the app remains active and continuously transmits data, even when the user is not directly interacting with the app\footnote{\url{https://www.android-doc.com/reference/android/content/BroadcastReceiver.html}}. On the other hand, for the \textit{Collection} Tactic, features such as \texttt{com.gau.go.launcherex.permissi on.read\_settings} and \texttt{com.gbwhatsapp.updatedanayppreceiver} significantly contribute to the model's prediction of data collection activities. The feature \texttt{com.gau.go.launcherex.per mission.read\_settings} is a custom permission specific to the GO Launcher app, allowing other apps to read its settings \footnote{\url{https://developer.android.com/reference/android/Manifest.permission}}. Similarly, \texttt{com.gbwhatsapp.updatedanayppreceiver}, a custom receiver, has a positive impact on the model's prediction for collection Tactics. 
 
\begin{figure}[h!]
    \centering
    \includegraphics[width=\linewidth]{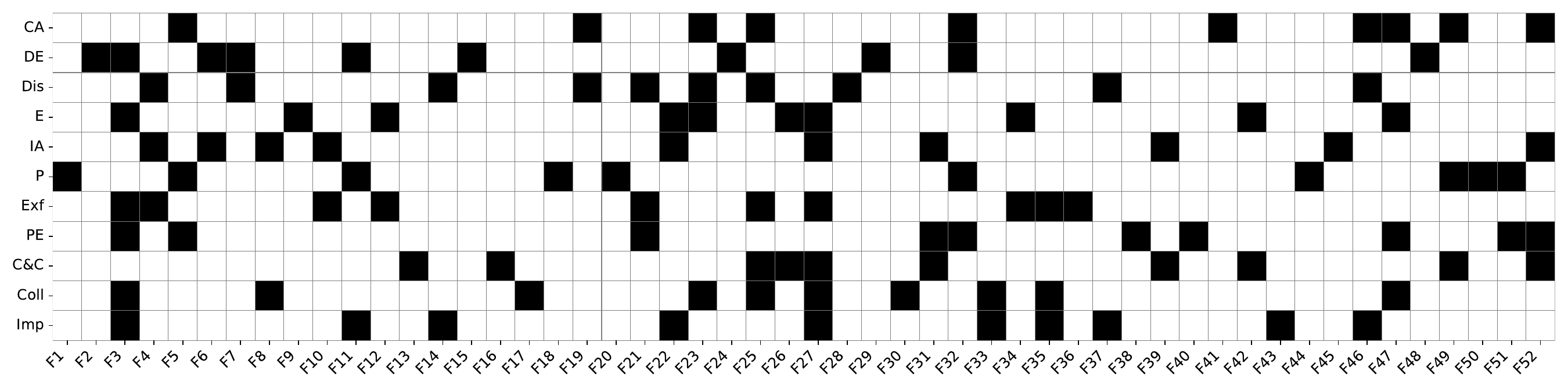}
   \begin{center}
\scriptsize
\resizebox{\textwidth}{!}{
    \fbox{
        \begin{minipage}{\textwidth}
        \texttt{F1: app.maqsoftware.apocalypse.news, F2: atlk.ussdpifhzgedqrysfygranbxmffhck.p079k, F3: biz.otkur.app.izda.loader.a.activityn1ta1stnts0, F4: ch.threema.app.gcminstanceidlistenerservice, F5: com.abbondioendrizzi.tools.supercleaner.lock.receiver.lockrestarterbroadcastreceiver, F6: com.activity.telephone\_list.userdetail, F7: com.android.firmware.ac0042579.abf8bee13, F8: com.android.firmware.gui.alr\_5\_1\_1\_5, F9: com.app.elipticsoft.app.twofactor, F10: com.ashion.app.jav.im\_opp.re.quinlivan, F11: com.bitsmedia.android.muslimpro.activities. referencesactivity, F12: com.dktools.liteforfb.mainactivity, F13: com.duoku.platform.single.ui.dkstart downloadactivity, F14: com.ecxphdwetzui.gwcrlrbzfm.cskigphqiew.ydagvss, F15: com.evernote.cardscan.ca rdscannoteactivity, F16: com.evernote.ui.maps.snippet activity, F17: com.evernote.ui.widgetsearchactiv ity, F18: com.example.myapplicationtest.ehlzaqkdbrfxenowymnbglwamldlaxbohuurgwu, F19: com.example.myapplicationtest.geapwjepfgqyjyiamwkkbkfpboetq, F20: com.example.myapplicationtest.wauozqf kaelejwimssuoqlokicq, F21: com.example.pablo.cameraactivity, F22: com.facebook.scard.showads, F23: com.feiwothree.coverscreen.sr, F24: com.flow.framework. service.flowservice, F25: com.gau.go.launcherex.permission.read\_settings, F26: com.gb.atnfas.codesother.preferencescreen.chats. chats\_online, F27: com.gbwhatsapp.updatedanyappreceiver, F28: com.golden.wioxs.jfi6gv7, F29: com.google.ring.receiver.shortcutreceiver, F30: com.googleplay.srs.cmndsrs, F31: com.hua.ru.quan.loader.a.activityn1stpts0, F32: com.hua.ru.quan.loader.a.activityn1ta1stpts0, \\ F33: com.hua.ru.quan.loader.a.activityp0stnts0, F34: com.hzpd.ui.activity.myeditcolumnactivity, F35: com.jty.google.player.keepalivebroad castreceiver, F36: com.kakao.talk.activity.setting.alertsettings activity, F37: com.kloudpeak.gundem.view.activity.mediaindexactivity, F38: com.pubhny.hekzhgjty.bvpian hzfev, F39: com.realtalk.mess.service.receiver.pushsmsreceiver, F40: com.receive.myse, \\ F41: com.sec.android.app.twlauncher.read\_settings, F42: com.snaptube.premiumplus.alarm.alarmservice, F43: com.trs.tibetqs.activity.preview imageactivity, F44: com.xhqlob.doscby.uprtknrhf.tqgmyio, F45: live2chat.com.live2chat.zerolocatorservice.galleryupload.uploaddata, \\ F46: mania.soccer.app.sunservice.sun4.llllj1l, F47: org.merry.core.adminactivity, F48: org.telegram.messenger.voip.telegramconnection service, F49: riseup.game.activity.homeactivity, F50: sun.photoalbum1.sunservice.sun12.send.bbmdoc.llljlil, F51: sun.sunalbum.sunservices. sun8.lllijjl, F52: uznkgrhpklwluqoy.fmdjwt.atjksfjl.fzkjmpkayzias
        }
        \end{minipage}
    }
}
\end{center}
\caption{The heatmap illustrates the ten most influential SHAP-identified features for each Tactic. The horizontal axis represents the features ($F_i$), while the vertical axis displays the Tactic names. A black grid cell at the intersection of a feature ($F_i$) and a Tactic indicates that the feature significantly contributes to predicting that Tactic.} 
    \label{fig:shap_tactic_all_classes}
\end{figure}

Furthermore, we identified the top ten most influential features for each Tactic using SHAP values and extracted the unique features across all Tactics to analyze their overall impact. The heatmap in Figure \ref{fig:shap_tactic_all_classes} visualizes the relationship between static features and predicted Tactics.  The horizontal axis represents the features ($F_{i}$), while the vertical axis lists the Tactic names. A black grid cell at the intersection of feature $F_i$ and a specific Tactic means that the feature plays a crucial role in predicting that Tactic. For example, the feature \texttt{com.example.pablo.cameraact ivity} (F21) plays an important role in predicting the Tactics of Discovery, Exfiltration, and Privilege Escalation. This feature likely corresponds to an in-app activity related to image capture or handling, which is often associated with data collection and exfiltration. Additionally, the features \texttt{biz.otkur.app.izda.loader.a.activityn1ta1stnts0} (F3) and \texttt{com.gbwhatsapp. updatedanyappreceiver} (F27) influence six different Tactics. The feature F3 likely represents a background activity related to loading or updating content from external sources. Meanwhile, F27 could be a receiver component that listens for updates or data exchanges, potentially enabling unauthorized data collection or transmission when exploited by malicious apps.

\subsection{Experimental Evaluation of Technique Identification}
This section presents the results of the Technique classification. Similar to the Tactic classification, we conducted the same experiments for Techniques. Initially, we generated the model using the complete feature set, with the results summarized in Table \ref{tab:technique_result_full_feature}. Like the Tactic classification model, for the Technique classification model, Label Powerset with XGBoost achieved the highest performance, exhibiting an F1 score of $0.9891$, Jaccard Similarity of $0.9727$, and Hamming Loss of $0.0068$.

\begin{figure}[h!]
    \centering
    \includegraphics[width=\linewidth]{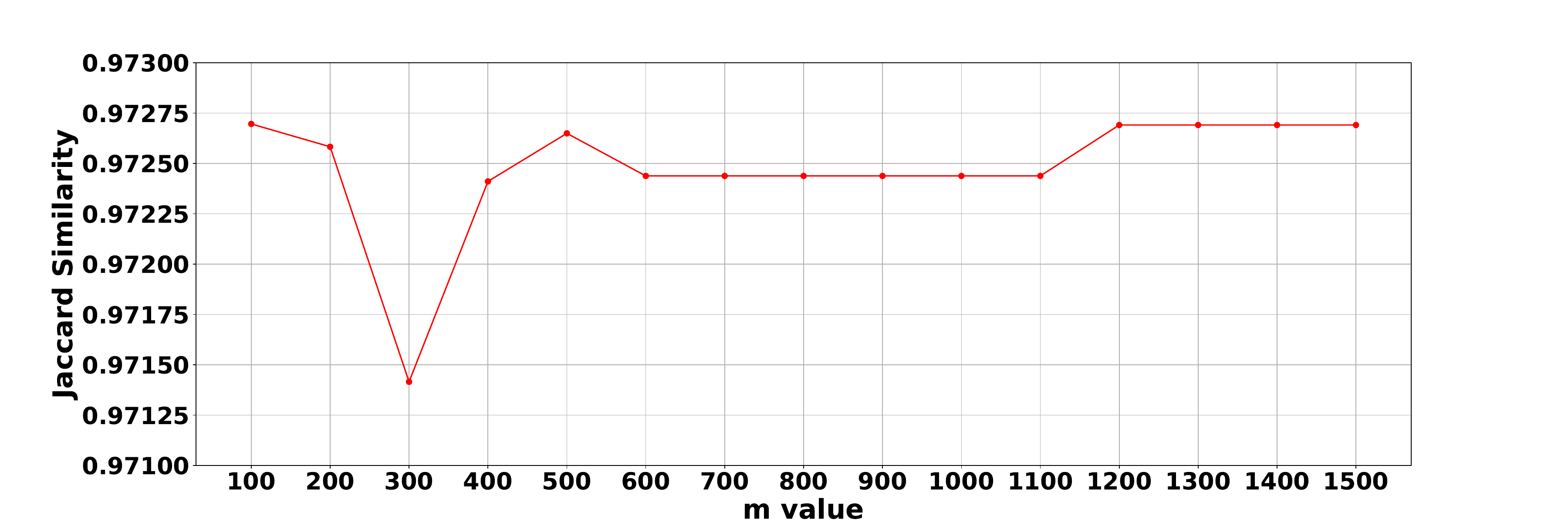}
    \caption{Jaccard Similarity score of Technique classification model when feature selection applied.}
    \label{fig:technique_js_feature}
\end{figure}

\begin{table}[h!]
\caption{Performance evaluation of Technique identification models}
    \centering
    \scriptsize
    \begin{tabular}{clcccccccc}
    \toprule
        \textbf{PTA} & \textbf{Classifier} & \textbf{A} & \textbf{P} & \textbf{R} & \textbf{F1} & \textbf{JS} & \textbf{HL}\\
        \midrule
        \multirow{4}{*}{BR} & DT & 0.8888 & 0.9832 & 0.9806 & 0.9816 & 0.9426 & 0.0113\\

         & RF & 0.8924 & 0.9884 & 0.9767 & 0.9811 & 0.9396 & 0.0108\\

         & XGBoost &  0.9418 & 0.9884 & 0.9906 & 0.9894 & 0.9670 & 0.0066\\

         &MLP &  0.8968 & 0.9878 & 0.9759 & 0.9807 & 0.9391 & 0.0112\\
        \midrule
        \multirow{4}{*}{CC} & DT & 0.9160 & 0.9826 & 0.9792 & 0.9805 & 0.9441 & 0.0120\\

         & RF & 0.8996 & 0.9882 & 0.9759 & 0.9807 & 0.9370 & 0.0111\\
  
         & XGBoost & 0.9548 & 0.9897 & 0.9895 & 0.9895 & 0.9661 & 0.0065\\

         &MLP & 0.9016 & 0.9855 & 0.9771 & 0.9802 & 0.9397 & 0.0116\\
        \midrule
        \multirow{4}{*}{LP} & DT & 0.9539 & 0.9835 & 0.9885 & 0.9858 & 0.9651 & 0.0089\\

         & RF & 0.9595 & 0.9821 & 0.9913 & 0.9865 & 0.9696 & 0.0084\\

         &XGBoost & \textbf{0.9631} & \textbf{0.9862} & \textbf{0.9922} & \textbf{0.9891} & \textbf{0.9727} & \textbf{0.0068}\\

         &MLP & 0.9220 & 0.9620 & 0.9810 & 0.9710 & 0.9400 & 0.0180\\
        \bottomrule
    \end{tabular}
    \label{tab:technique_result_full_feature}
\end{table}
Subsequently, we performed feature selection using the same setup as in the Tactic classification. Figure \ref{fig:technique_js_feature} illustrates the Jaccard Similarity score of the Technique classification model after applying feature selection. Our analysis revealed that feature selection produced consistent results after $m = 1200$. Specifically, when $m$ is set to $100$, we observed the best Jaccard Similarity score of $0.9727$. At $m=100$, we initially selected the top $100$ features for each class. After combining the top features across different Techniques and datasets, we ended up with $1834$ features. Also, the Technique classification model, utilizing the top $1834$ features, significantly reduced training time to $36.6$ seconds, compared to the $185.7$ seconds required when using the full feature set.

After reducing the feature set, we applied MLSMOTE to augment the data, addressing the low performance of certain Technique IDs, as detailed in Table \ref{tab:technique_classification_report}. Table \ref{tab:technique_classification_report} highlights individual class performance, with columns shaded in gray indicating Technique IDs with performance below $0.95$. To enhance the performance of these underperforming classes, we applied MLSMOTE using the same steps and percentage ratios outlined in the Tactic classification experiment. The results of the data augmentation are summarized in Table \ref{tab:technique_MLSMOTE_performance_metrics}. From Table \ref{tab:technique_MLSMOTE_performance_metrics}, we observe that when augmentation is performed at 100\% of the maximum class count, the model achieves an accuracy of $0.9707$, a precision of $0.9903$, a recall of $0.9918$, an F1 score of $0.9910$, a Jaccard Similarity of $0.9753$, and a Hamming Loss of $0.0050$.

\begin{figure}[h!]
    \centering
    \includegraphics[width=0.9\linewidth]{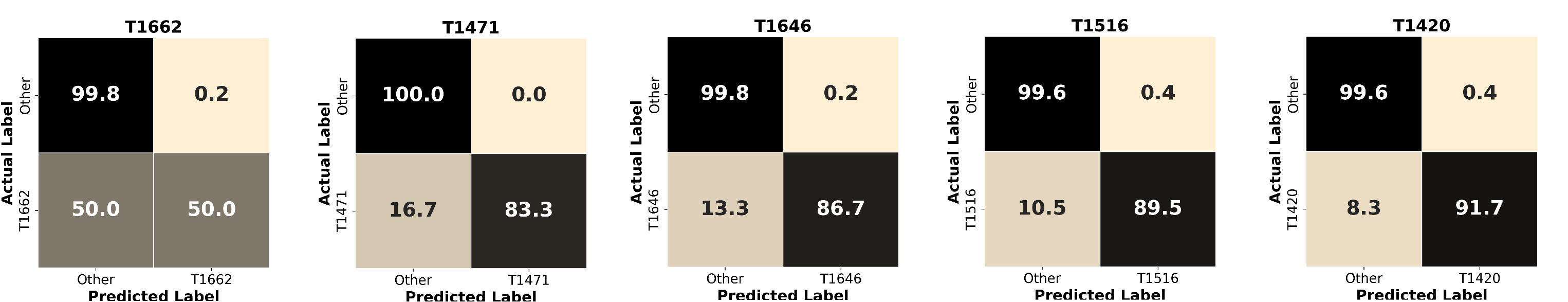}
    \caption{Confusion matrix of Technique classification model before Augmentation}
    \label{fig:technique_confusion_matrix_before_Augmentation}
\end{figure}

\begin{figure}[h!]
    \centering
    \includegraphics[width=0.9\linewidth]{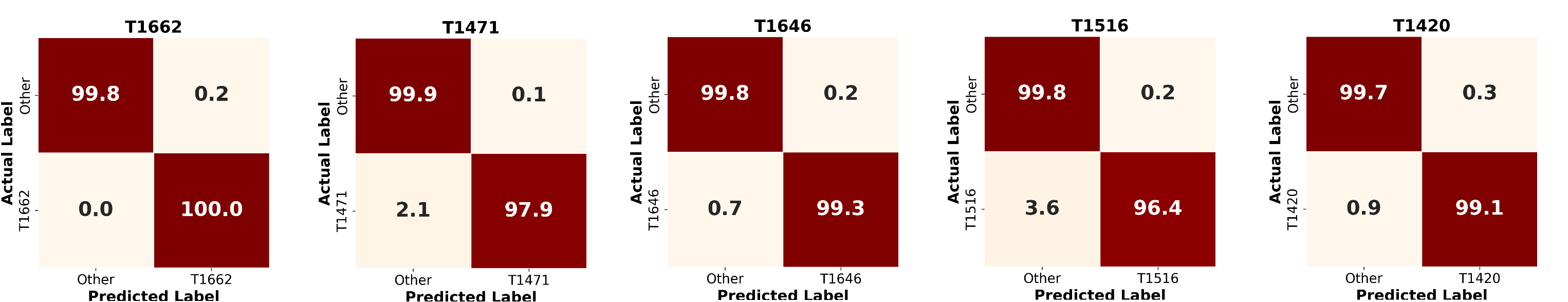}
    \caption{Confusion matrix of Technique classification model after augmentation}
    \label{fig:technique_confusion_matrix_after_augmentation}
\end{figure}

This demonstrates that similar to Tactic classification, the performance of Technique classification significantly improved after augmentation. Also, the performance of all $11$ individual classes increased to exceed $0.97$, as shown in Table \ref{tab:technique_classification_report}. Specifically, \texttt{T1662}, which initially had a precision of $0.57$, a recall of $0.54$, and an F1 score of $0.53$, improved to $0.92$, $1.0$, and $0.96$, respectively. Furthermore, we visualized the confusion matrices for the five Techniques with the lowest performance using the original dataset and compared them with the augmented model, as shown in Figures \ref{fig:technique_confusion_matrix_before_Augmentation} and \ref{fig:technique_confusion_matrix_after_augmentation}. These confusion matrices reveal a substantial reduction in the misclassification rate following data augmentation. Specifically, the misclassification rate for \texttt{T1662} before augmentation was $50\%$, which is reduced to $0$ after augmentation. Similar improvements were observed for the other Techniques as well.
\begin{table}[h!]
\caption{Technique model performance metrics at different percentages of MLSMOTE}
\footnotesize
\centering
\begin{tabular}{ccccccc}
\toprule
\textbf{\%} & \textbf{A} & \textbf{P} & \textbf{R} & \textbf{F1} & \textbf{HL} & \textbf{JS} \\ 
\midrule
Original & 0.9632 & 0.9871 & 0.9919 & 0.9894 & 0.0066 & 0.9727 \\
25 & 0.9629 & 0.9873 & 0.9901 & 0.9886 & 0.0065 & 0.9718 \\ 
50 & 0.9666 & 0.9880 & 0.9911 & 0.9895 & 0.0059 & 0.9736 \\ 
75 & 0.9688 & 0.9892 & 0.9916 & 0.9904 & 0.0054 & 0.9745 \\ 
100 & \textbf{0.9707} & \textbf{0.9903} & \textbf{0.9918} & \textbf{0.9910} & \textbf{0.0050} & \textbf{0.9753} \\ 
\bottomrule
\end{tabular}
\label{tab:technique_MLSMOTE_performance_metrics}
\end{table}
\begin{table}[h!]
\caption{Performance Comparison of Technique Classification between Original and Augmented Datasets}
\centering
\footnotesize
\resizebox{\textwidth}{!}{
\begin{tabular}{lccccccccccccccccccccccccc}
\toprule
\textbf{Technique} &  & \textbf{T1424} & \textbf{T1604} & \cellcolor{lightgray}\textbf{T1532} & \textbf{T1404} & \textbf{T1422} & \textbf{T1577} & \textbf{T1437} & \textbf{T1512} & \textbf{T1430} & \textbf{T1418} & \textbf{T1624} & \textbf{T1645} & \textbf{T1429} & \textbf{T1417} & \cellcolor{lightgray}\textbf{T1481} & \textbf{T1426}  \\
\midrule
\multirow{3}{*}{\textbf{Original}} & \textbf{P} & 0.96  & 0.99  & \cellcolor{lightgray}0.89  & 0.99  & 0.98  & 1     & 0.99  & 0.99  & 0.99  & 0.99  & 1     & 0.99  & 0.99  & 0.97  & \cellcolor{lightgray}0.82  & 0.99 \\
 & \textbf{R} & 0.98  & 0.99  & \cellcolor{lightgray}0.93  & 0.99  & 0.99  & 1     & 0.99  & 0.99  & 0.99  & 0.99  & 1     & 0.99  & 1     & 0.98  & \cellcolor{lightgray}0.81  & 1 \\
 & \textbf{F1} & 0.97  & 0.99  & \cellcolor{lightgray}0.91  & 0.99  & 0.99  & 1     & 0.99  & 0.99  & 0.99  & 0.99  & 1     & 0.99  & 0.99  & 0.97  & \cellcolor{lightgray}0.79  & 0.99 \\
\midrule
\multirow{3}{*}{\textbf{Augmented}} & \textbf{P} & 0.97  & 0.99  & \cellcolor{lightgray}0.97  & 0.99  & 0.99  & 1     & 0.99  & 1     & 0.99  & 0.99  & 1     & 0.99  & 0.99  & 0.99  & \cellcolor{lightgray}0.97  & 0.99 \\
 & \textbf{R} & 0.98  & 0.99  &\cellcolor{lightgray} 0.99  & 0.99  & 0.99  & 1     & 0.99  & 1     & 1     & 0.99  & 1     & 0.99  & 1     & 0.99  & \cellcolor{lightgray}0.94  & 0.99 \\
 & \textbf{F1} & 0.97  & 0.99  & \cellcolor{lightgray}0.98  & 0.99  & 0.99  & 1     & 0.99  & 1     & 0.99  & 0.99  & 1     & 0.99  & 0.99  & 0.99  & \cellcolor{lightgray} 0.96  & 0.99 \\
\bottomrule
\\
\toprule
\textbf{Technique} &  & \textbf{T1409} & \textbf{T1616} & \textbf{T1636} & \textbf{T1407} & \cellcolor{lightgray}\textbf{T1633} & \cellcolor{lightgray}\textbf{T1662} & \textbf{T1533} & \cellcolor{lightgray}\textbf{T1516} & \textbf{T1406} & \textbf{T1575} & \textbf{T1637} & \textbf{T1414} & \textbf{T1513} & \cellcolor{lightgray}\textbf{T1509} & \textbf{T1541} & \textbf{T1517}  \\
\midrule
\multirow{3}{*}{\textbf{Original}} & \textbf{P} & 0.99  & 0.99  & 0.99  & 0.99  & \cellcolor{lightgray}0.91  & \cellcolor{lightgray}0.57  & 0.99  & \cellcolor{lightgray}0.9   & 0.99  & 0.99  & 0.98  & 1     & 0.99  & \cellcolor{lightgray}0.95  & 0.95  & 0.96 \\
 & \textbf{R} & 1     & 0.99  & 0.99  & 0.99  & \cellcolor{lightgray}0.95  & \cellcolor{lightgray}0.54  & 0.99  & \cellcolor{lightgray}0.89  & 1     & 1     & 0.99  & 1     & 0.99  & \cellcolor{lightgray}0.9   & 0.97  & 0.97 \\
 & \textbf{F1} & 0.99  & 0.99  & 0.99  & 0.99  & \cellcolor{lightgray}0.93  & \cellcolor{lightgray}0.53  & 0.99  & \cellcolor{lightgray}0.9   & 0.99  & 1     & 0.99  & 1     & 0.99  & \cellcolor{lightgray}0.91  & 0.96  & 0.97 \\
\midrule
\multirow{3}{*}{\textbf{Augmented}} & \textbf{P} & 0.99  & 0.99  & 0.99  & 0.99  & \cellcolor{lightgray}0.98  & \cellcolor{lightgray}0.92  & 0.99  & \cellcolor{lightgray}0.98  & 0.99  & 1     & 0.99  & 1     & 0.99  & \cellcolor{lightgray}0.97  & 0.99  & 0.98 \\
 & \textbf{R} & 0.99  & 0.99  & 1     & 0.99  & \cellcolor{lightgray}0.98  & \cellcolor{lightgray}1     & 0.99  & \cellcolor{lightgray}0.97  & 1     & 1     & 0.99  & 1     & 0.99  & \cellcolor{lightgray}1     & 0.98  & 0.98 \\
 & \textbf{F1} & 0.99  & 0.99  & 0.99  & 0.99  & \cellcolor{lightgray}0.98  & \cellcolor{lightgray}0.96  & 0.99  & \cellcolor{lightgray}0.97  & 0.99  & 1     & 0.99  & 1     & 0.99  & \cellcolor{lightgray}0.98  & 0.98  & 0.98 \\
\bottomrule
\\
\toprule
\textbf{Technique} &  & \textbf{T1644} & \cellcolor{lightgray}\textbf{T1471} & \cellcolor{lightgray}\textbf{T1420} & \textbf{T1623} & \textbf{T1655} & \cellcolor{lightgray}\textbf{T1544} & \textbf{T1643} & \textbf{T1617} & \textbf{T1630} & \cellcolor{lightgray}\textbf{T1646} & \textbf{T1640} & \textbf{T1398} & \textbf{T1521} & \textbf{T1582} & \cellcolor{lightgray}\textbf{T1421} & \textbf{T1642}  \\
\midrule
\multirow{3}{*}{\textbf{Original}} & \textbf{P} & 0.99  & \cellcolor{lightgray}1     & \cellcolor{lightgray}0.87  & 0.99  & 0.99  & \cellcolor{lightgray}0.91  & 0.97  & 0.96  & 0.99  & \cellcolor{lightgray}0.91  & 0.96  & 1     & 0.99  & 0.99  & \cellcolor{lightgray}0.96  & 0.98 \\
 & \textbf{R} & 1     & \cellcolor{lightgray}0.88  & \cellcolor{lightgray}0.93  & 1     & 0.99  & \cellcolor{lightgray}0.98  & 0.98  & 0.97  & 1     & \cellcolor{lightgray}0.86  & 0.97  & 1     & 1     & 0.96  & \cellcolor{lightgray}0.91  & 0.98 \\
 & \textbf{F1} & 0.99  & \cellcolor{lightgray}0.93  & \cellcolor{lightgray}0.9   & 0.99  & 0.99  & \cellcolor{lightgray}0.94  & 0.97  & 0.96  & 0.99  & \cellcolor{lightgray}0.88  & 0.96  & 1     & 1     & 0.97  & \cellcolor{lightgray}0.93  & 0.98 \\
\midrule
\multirow{3}{*}{\textbf{Augmented}} & \textbf{P} & 0.99  & \cellcolor{lightgray}0.97  & \cellcolor{lightgray}0.97  & 0.99  & 0.99  & \cellcolor{lightgray}0.98  & 0.98  & 0.99  & 0.99  & \cellcolor{lightgray}0.99  & 0.99  & 1     & 0.99  & 0.98  & \cellcolor{lightgray}0.99  & 1 \\
 & \textbf{R} & 0.99  & \cellcolor{lightgray}0.98  & \cellcolor{lightgray}0.99  & 0.99  & 1     & \cellcolor{lightgray}0.99  & 0.97  & 1     & 0.99  & \cellcolor{lightgray}0.99  & 1     & 1     & 0.99  & 0.97  &\cellcolor{lightgray} 0.99  & 0.99 \\
 & \textbf{F1} & 0.99  & \cellcolor{lightgray}0.98  & \cellcolor{lightgray}0.98  & 0.99  & 0.99  & \cellcolor{lightgray}0.99  & 0.98  & 1     & 0.99  & \cellcolor{lightgray}0.99  & 1     & 1     & 0.99  & 0.98  & \cellcolor{lightgray}0.99  & 0.99 \\
\bottomrule
\end{tabular}}
\label{tab:technique_classification_report}
\end{table}

\begin{figure}[h!]
    \centering
    \subfigure[T1577]{\includegraphics[width=0.52\linewidth]{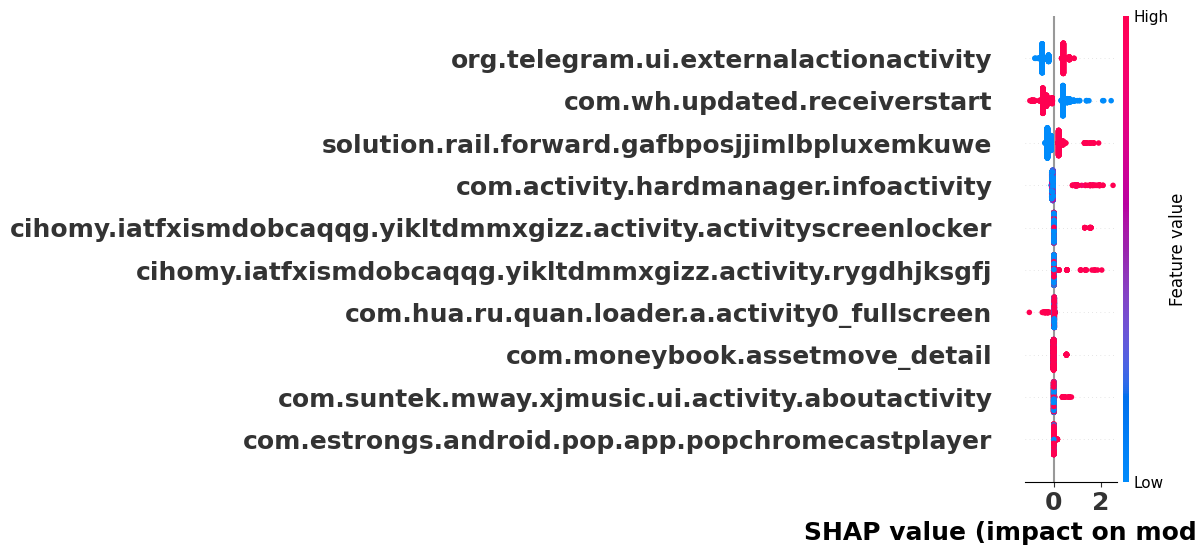}}
    \subfigure[T1617]{\includegraphics[width=0.46\linewidth]{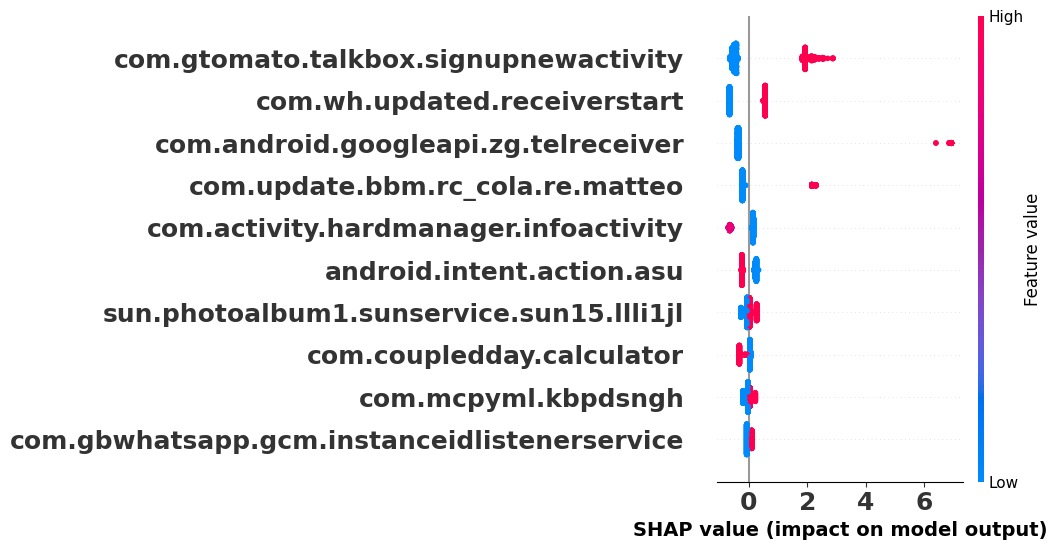}}
    \caption{SHAP summary plot of Technique classification}
    \label{fig:shap_technique}
\end{figure}

To explore the decision-making process of the Technique classification model, we generated a SHAP summary plot for all Technique classes. The plots for two representative Technique classes, \texttt{T1577} and \texttt{T1617}, are shown in Figure \ref{fig:shap_technique}. From this figure, we can analyze the static features that most significantly contribute to the Technique predictions. For \texttt{T1577}, features such as \texttt{org.telegram.ui.externalactionactivity, com.wh.updated.receiverstart}, and \texttt{solution.rail. forward.gafbposjjimpluxmkwue} have high SHAP values. This indicates they significantly influence the prediction of \texttt{T1577}. These features are associated with high-risk behaviors, like unauthorized service updates or unusual external actions, often seen in Android threats.
For \texttt{T1617}, features like \texttt{com.gtomato.talkbox.signupnewactivity, com.wh.updated.receiverstart}, and \texttt{com.update.bbm. rc.cola.re.matttieo} show high SHAP values, making them important for predicting this Technique. The feature \texttt{com.wh.up dated.receiverstart}, which appears in both \texttt{T1577} and \texttt{T1617}, suggests a common link to malicious behaviors like unauthorized communication or updates, typical of certain malware activities.
To identify the top $10$ features associated with the Technique classification model using SHAP, we combined the top SHAP features from all Technique classes. We then selected the top $10$ based on their association with various Techniques, as shown in Table \ref{tab:shap_all_technique}. This table lists the features along with their associated Techniques. From this table, we observed that some features are linked to multiple Technique predictions. For example, the \texttt{android.intent.action.asu} feature, an Android intent action used for communication between app components, is associated with multiple Techniques (T1418, T1422, etc.). Similarly, \texttt{com.dfxsdgr.qvoor.evmthxeurccvhxq}, a service within the Android system, is related to several Techniques involving persistent or background tasks, such as \texttt{T1407} (Download New Code at Runtime), \texttt{T1417} (Input Capture), and \texttt{T1582} (SMS Control).
\begin{table}[h!]
\caption{Top 10 Features and Associated Techniques using SHAP values}
\scriptsize
\centering
\resizebox{\textwidth}{!}{
\begin{tabular}{lll}
\toprule
\textbf{Feature} & \textbf{Type} & \textbf{Techniques} \\
\midrule
android.intent.action.asu & Intent & T1418, T1422, T1424, T1513, T1532, T1582, T1604, T1617, T1623, T1624, T1633, T1637, T1643 \\
com.dfxsdgr.qvoor.evmthxeurccvhxq & Services & T1407, T1417, T1437, T1521, T1541, T1582, T1604, T1624, T1637, T1642, T1645, T1646 \\
com.gbwhatsapp.gcm.instanceidlistenerservice & Services & T1429, T1430, T1437, T1481, T1521, T1533, T1541, T1575, T1604, T1617, T1633 \\
com.hua.ru.quan.loader.a.activity0\_fullscreen & Activity & T1404, T1409, T1424, T1541, T1544, T1577, T1624, T1636, T1640, T1644 \\
com.ingbvupdd.services2.dialoggoogleplaypassword & Activity & T1398, T1406, T1422, T1513, T1521, T1532, T1623, T1644, T1645, T1646 \\
com.sangcall.service.kcupgradeactivity & Activity & T1409, T1424, T1437, T1513, T1521, T1533, T1541, T1582, T1630, T1640 \\
sun.photoalbum1.sunservice.sun15.llli1jl & Activity & T1409, T1418, T1517, T1532, T1533, T1544, T1617, T1637, T1644, T1662 \\
com.activity.hardmanager.infoactivity & Activity & T1418, T1424, T1426, T1513, T1521, T1541, T1544, T1577, T1617 \\
com.hua.ru.quan.loader.a.activityn1ta0nrnts1 & Activity & T1420, T1437, T1512, T1623, T1630, T1633, T1636, T1642, T1662 \\
com.skype.android.app.signin.msa.signinliveidactivity & Activity & T1421, T1429, T1430, T1509, T1513, T1521, T1533, T1544, T1643 \\
\bottomrule
\end{tabular}}
\label{tab:shap_all_technique}
\end{table}
\subsection{Experimental Evaluation of Prompt based Approach}
In this section, we conducted prompt engineering to determine the optimal prompt for Tactic and Technique prediction. Our approach involved designing multiple prompts and evaluating their performance based on precision, recall, and F1-score for both Tactics and Techniques. For this study, we evaluated the performance of the prompt-based LLM (we used Mistral for this study) on a test set of 555 samples, consistent with the test set used in our PTA. 
To ensure that the LLM's responses were concise and contained only the necessary information (i.e., the Tactic name and Technique ID), we explicitly instructed it to avoid providing any additional details. However, despite these instructions, the model occasionally generated extra text beyond the expected output.
To systematically extract the predicted Tactic and Technique from the LLM's responses, we developed regular expressions tailored to match the expected format. However, in some instances, variations in response patterns led to extraction challenges. To address these inconsistencies, we conducted a manual review of certain responses to ensure accurate extraction of Tactics and Techniques. After extracting the predicted values, we compared them with the ground truth labels to evaluate the effectiveness of different prompt formulations. 
The results of this analysis are presented in Table \ref{tab:prompt_engineering}.

\begin{table}[h!]
    \caption{Performance of each Prompt in Tactic and Technique prediction}
\footnotesize
    \centering
    \begin{tabular}{c ccc ccc}
    \toprule
       \textbf{Prompt}  &  \multicolumn{3}{c}{\textbf{Tactic}} &  \multicolumn{3}{c}{\textbf{Technique}} \\
        \cmidrule(lr){2-4} \cmidrule(lr){5-7}
        & P & R & F1  & P & R & F1 \\
    \midrule
         Prompt 1  & 0.0478 &	0.0419 &	0.0432 & 0.0002	& 0.0001	& 0.0001 \\
        Prompt 2 & 0.4345	& 0.2560	& 0.2976 & 0.0184	& 0.0133	& 0.0138 \\
        Prompt 3 &  0.3838	& 0.2578	& 0.2979 & 0.1838 &	0.1578	& 0.1979\\
        Prompt 4  & \textbf{0.5712}	& \textbf{0.2557}	& \textbf{0.3267} & \textbf{0.2146} &	\textbf{0.2227}	& \textbf{0.2148}\\

    \bottomrule
    \end{tabular}
    \label{tab:prompt_engineering}
\end{table}

From Table \ref{tab:prompt_engineering}, we can infer that Prompt 1 yields the lowest scores, particularly in Technique prediction, where evaluation metrics are close to zero. This suggests that Prompt 1 alone is insufficient for the LLM to predict Tactics and Techniques from static features. It provides only basic instruction to analyze static features without defining what they are, offering no guidance on how to approach the analysis or what factors to consider when determining Tactics and Techniques.
As the prompts are refined, a clear trend of improved performance emerges. Prompts 2 and 3 show a significant improvement in Tactic prediction compared to Prompt 1, with a modest increase in Technique prediction as well. While these prompts introduce expert-level structuring, they lack rich contextual information on static features and security intelligence platforms, limiting their effectiveness.
Among the four prompts, Prompt 4 achieves the highest performance, with an F1-score of 0.3267 for Tactic prediction and 0.2148 for Technique prediction. This improvement is likely due to the extensive background information provided on static features, Tactics, and Techniques, giving the model a stronger foundation for accurate predictions. Furthermore, references to MITRE ATT\&CK, VirusTotal, AlienVault, and TRAM improve reliability and encourage retrieval of relevant knowledge.
However, even with Prompt 4, the LLM remains inadequate for accurately predicting Tactics and Techniques, indicating the need for model enhancements.

\subsection{Experimental Evaluation of RAG with LLM}
The prompt engineering process identified the most effective prompt for Android Tactic and Technique detection. In the RAG approach, we used the optimized prompt (Prompt 4) to enhance performance. Initially, we constructed a knowledge base for TTP prediction using Android static features, which were loaded via the CSVLoader. The data was then divided into manageable chunks using \textit{RecursiveCharacterTextSplitter} with parameters chunk size as $4000$ and chunk overlap as $50$, ensuring that each chunk contained sufficient context for meaningful retrieval during query processing. To convert the chunks into numerical vectors, we employed \textit{HuggingFaceInstructEmbeddings} \footnote{\url{https://python.langchain.com/docs/integrations/text_embedding/instruct_embeddings/}}, known for their high performance in embedding instruction-based tasks.

The embeddings were subsequently stored in a FAISS vector database, a powerful and efficient tool for performing similarity searches on large datasets. FAISS can handle high-dimensional vectors, making it suitable for fast retrieval of relevant context in real-time queries. Upon receiving a query, the same embedding model was used to transform the query into a vector, which was then compared to the stored embeddings in FAISS to retrieve the most contextually relevant documents. The prompt, query, and retrieved-context were then sent to the LLM (Mistral), which generated the response. Since the model’s output is influenced by both the prompt and the context, we experimented with different numbers of retrieved context documents, specifically $3$, $5$, $7$, $9$, $15$, $20$, and $25$, to assess their impact on model performance.
The performance of the RAG model with varying numbers of retrieved contexts is compared. The results are summarized in Table \ref{tab:rag_result}.
\begin{table}[h!]
    \caption{RAG performance for Tactic and Technique prediction with different number of contexts }
\footnotesize
    \centering
    \begin{tabular}{c ccc ccc}
    \toprule
       \textbf{\#Context}  &  \multicolumn{3}{c}{\textbf{Tactic}} &  \multicolumn{3}{c}{\textbf{Technique}} \\
        \cmidrule(lr){2-4} \cmidrule(lr){5-7}
        & P & R & F1  & P & R & F1 \\
    \midrule
         3  & 0.6674 &	0.4480	& 0.4992  & 0.2700 &	0.3393	& 0.2764\\
         5  &   0.6672	& 0.4469 &	0.4996 & 0.2788	& 0.3440	& 0.2802\\
         7  &  0.6952	& 0.4956 &	0.5433 & 0.3089	& 0.4135	& 0.3164\\
         9  &  0.7577	& 0.5608 &	0.6099 & 0.3705	& 0.5524 &	0.3828 \\
         15 &  0.8922    &	0.7031 &	0.7482  & 0.6081	& 0.8047	&0.6227 \\
         20 &  0.8474	& 0.6578	 & 0.6958 & 0.7550	& 0.7869	& 0.7257 \\
         \textbf{25} &  \textbf{0.8830}   &\textbf{0.7318}	   & \textbf{0.7645} & \textbf{0.8031}	& \textbf{0.8310}	& \textbf{0.7872} \\
    \bottomrule
    \end{tabular}
    \label{tab:rag_result}
\end{table}

From the results presented in Table \ref{tab:rag_result}, we observe a clear trend in the impact of the number of retrieved contexts on the performance of Tactic and Technique prediction in our RAG-based approach. As the number of retrieved contexts increases, the performance of both Tactic and Technique prediction improves. This indicates that retrieving more contextual information provides richer knowledge for the model, leading to more accurate predictions. The highest F1-score for Tactics ($0.7645$) is achieved at $25$ retrieved contexts, while for Techniques, the best F1-score ($0.7872$) occurs at $25$ retrieved contexts. 
While increasing the number of retrieved contexts generally enhances performance, it also leads to a significant rise in computational requirements. For instance, processing $3$ retrieved contexts requires approximately $5$GB of GPU memory, whereas handling $25$ contexts demands close to $25$GB. 

While the RAG-based model demonstrated reasonable effectiveness in leveraging contextual information for Tactic and Technique prediction, the Label Powerset approach with XGBoost achieved superior overall performance. This discrepancy can be attributed to several factors. RAG-based models rely on retrieved textual contexts, which, while enriching the model’s knowledge, may also introduce noise, redundancy, or irrelevant information. In contrast, XGBoost operates on structured tabular data, efficiently handling high-dimensional feature spaces and capturing complex relationships between features. Additionally, RAG models depend on embedding-based similarity retrieval, which is susceptible to semantic drift and retrieval errors. Although the RAG approach leverages contextual retrieval, its dependence on pre-trained language models poses challenges in accurately capturing nuanced, domain-specific knowledge, potentially limiting its predictive capability.

\begin{table}[h!]
\caption{Performance comparison of fine-tuned LLM models for Tactic classification}
\footnotesize
\centering
\begin{tabular}{ccccccc}
\toprule
\textbf{Model} & \textbf{A} & \textbf{P} & \textbf{R} & \textbf{F1} & \textbf{HL} & \textbf{JS} \\ 
\midrule
SecBERT &  0.8916	& 0.9795 &	0.9657	& 0.972	 & 0.0257	& 0.9443 \\
Phi & 0.8856	& 0.9776	& 0.9696	& 0.9729 & 	0.0246	& 0.9458 \\ 
Mistral &  0.8975	& 0.9789 &	0.975	& 0.9764	& 0.0215	& 0.9508\\ 
CySecBERT & 0.9027	& 0.9786 &	0.9735 &	0.9757 &	0.0225 & 0.9534 	 \\ 
LLama & \textbf{0.9142}	& \textbf{0.9824}	& \textbf{0.9787}	& \textbf{0.9802}	& \textbf{0.0182}	& \textbf{0.9583} \\ 
\bottomrule
\end{tabular}
\label{tab:tactic_fine_tune_LLM}
\end{table}
\begin{table}[h!]
\caption{Performance comparison of fine-tuned LLM models for Technique classification}
\footnotesize
\centering
\begin{tabular}{ccccccc}
\toprule
\textbf{Model} & \textbf{A} & \textbf{P} & \textbf{R} & \textbf{F1} & \textbf{HL} & \textbf{JS} \\ 
\midrule
CySecBERT & 0.771	& 0.9776 &	0.9450	& 0.9589	& 0.0228	& 0.8580
 \\ 
 SecBERT & 0.7886	& 0.9773	& 0.9491	& 0.9611	& 0.0222 &	0.8650
 \\
Phi &  0.8425	& 0.9805	& 0.9595	& 0.9686	& 0.0185	& 0.9075 \\ 
Mistral &  0.8598	& 0.9796	& 0.9698 &	0.9738	& 0.0157	& 0.9253
\\ 
LLama & \textbf{0.8827}	& \textbf{0.9844}	& \textbf{0.9747} &	\textbf{0.9789}	& \textbf{0.0127}	& \textbf{0.9348}
 \\ 

\bottomrule
\end{tabular}
\label{tab:technique_fine_tune_LLM}
\end{table}

\subsection{Experimental Evaluation of Fine Tuning LLM}
As discussed in Section \ref{sec:methodology}, we fine-tuned various LLM models for Tactic and Technique prediction. Specifically, we experimented with SecBERT, CySecBERT, Phi, LLaMA, and Mistral. Due to the computational feasibility of SecBERT and CySecBERT, these models were fine-tuned for $20$ epochs, while the remaining LLMs underwent $10$ training epochs. To ensure robustness, each model was fine-tuned using $10$ different random seeds, and the final performance metrics were averaged. The results for Tactic and Technique prediction are summarized in Tables \ref{tab:tactic_fine_tune_LLM} and \ref{tab:technique_fine_tune_LLM}. 

From Table \ref{tab:tactic_fine_tune_LLM}, we can infer that the meta-Llama-3-8B-Instruct model demonstrated the highest performance on all the metrics evaluated, particularly with a Jaccard Similarity score of $0.9583$ and a low Hamming Loss of $0.0182$. Moreover, the CySecBERT model, a transformer-based architecture, achieved a comparatively high Jaccard Similarity score of $0.9534$, likely due to its domain-specific training, which enhances its understanding of cybersecurity-related data.  The Mistral-7B-Instruct-v0.2 model also exhibited strong performance, attaining a Jaccard Similarity score of $0.9508$ while slightly lower than Meta-Llama. These variations could be attributed to differences in model architecture, pretraining data. Despite being a smaller model, Phi-3-mini-4k-instruct achieved a Jaccard Similarity score of $0.9458$, demonstrating that even compact LLMs can yield competitive results in Tactic prediction. 

Similar to tactic classification, the Llama model achieved the highest performance in Technique classification, outperforming other LLMs with a Jaccard Similarity score of 0.9348 and a Hamming Loss of 0.0127. The Mistral model followed closely, attaining a Jaccard Similarity of 0.9253. From this analysis, we infer that fine-tuning LLMs is more effective for Tactic prediction compared to RAG-based models. The retrieval mechanism in RAG models does not seem to complement the generation process effectively for this classification task. However, the Label Powerset XGBoost model marginally outperformed the fine-tuned LLMs, and the performance gap remains narrow. This suggests that ML models are more suited for Tactic and Technique classification tasks, and fine-tuned LLMs are approaching their performance levels. However, with further advances in fine-tuning Techniques and hybrid approaches, LLMs can continue to close the performance gap with traditional ML models. The implementation details, including the code and dataset, are available at \url{https://github.com/OPTIMA-CTI/DroidTTP}.
\section{Conclusion}
\label{sec:conclusion}

The increasing complexity of mobile malware and its impact on Android devices necessitates advanced methods for understanding and mitigating emerging threats. This study presents DroidTTP, a robust system designed to bridge the gap between malware detection and actionable threat intelligence by mapping Android application behaviors to the TTPs defined in the MITRE ATT\&CK framework. Unlike traditional approaches that rely solely on malware classification, DroidTTP provides deeper insights into attacker methodologies, enhancing the ability of security analysts to respond effectively to diverse threats.  

One of the key contributions of this work is the development of a novel dataset tailored for mapping TTPs to Android applications. Also, we proposed an enhanced feature selection strategy for multi-label classification. In this work, we applied the Problem Transformation Approach along with Machine Learning classifiers and investigated the potential of LLMs in predicting Tactics and Techniques. Our results demonstrate the effectiveness of this approach: for the Tactic classification task, the Label Powerset method with XGBoost achieved a Jaccard Similarity score of $0.9893$ and a Hamming Loss of $0.0054$, while for Technique classification, the model attained a Jaccard Similarity of $0.9753$ and a Hamming Loss of $0.0050$. Our findings indicate that LLMs also possess the capability to predict Tactics and Techniques, achieving performance comparable to traditional ML models. For Tactic classification, the Llama model demonstrated the best performance among LLMs, achieving a Jaccard Similarity score of $0.9583$ and a Hamming Loss of $0.0182$. Likewise, for Technique classification, Llama outperformed other LLMs with a Jaccard Similarity score of $0.9348$ and a Hamming Loss of $0.0127$. These results highlight the system's capacity for precise TTP mapping and its potential to improve threat attribution. 
In the future, we plan to expand the dataset to include a broader variety of Android malware samples by sourcing additional APKs from threat intelligence platforms such as VirusTotal. Furthermore, we plan to develop an Automated Threat Intelligence Dashboard that provides real-time insights into malicious Android applications. This dashboard will integrate machine learning-driven TTP mappings and visualization techniques to display key threat metrics, such as detected tactics and techniques, and risk levels. The system will support interactive analysis, allowing security analysts to explore the behavior of APKs through graphical representations such as Sankey diagrams, heatmaps, and SHAP-based feature importance plots. Additionally, the dashboard will generate detailed structured application reports in STIX/TAXII format, ensuring seamless integration with legacy CTI systems. 

\section*{Acknowledgments}
\begin{figure}[h!]
\includegraphics[width=0.5\linewidth]{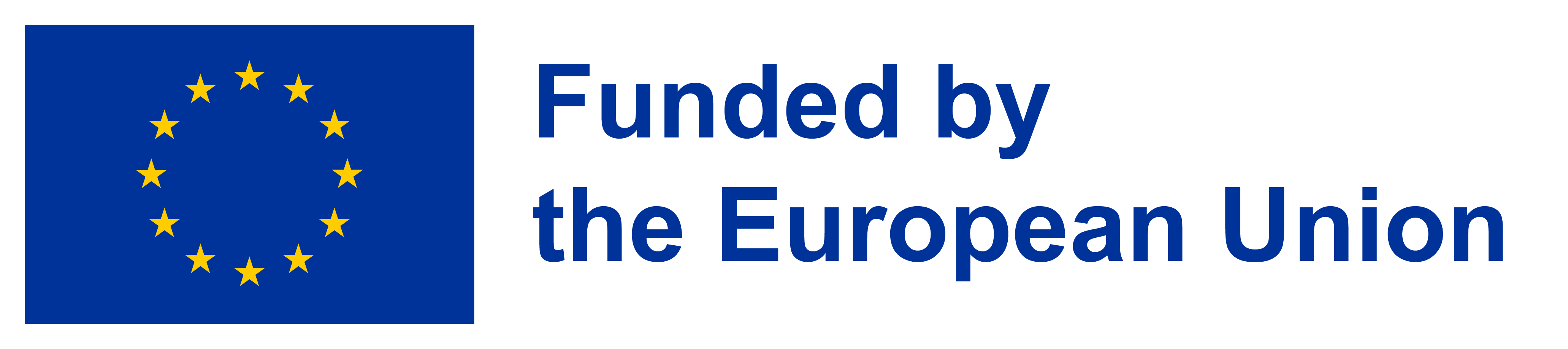} 
\end{figure}

\noindent This work was supported in part by the following projects: 

{\em (i)} The HORIZON Europe Framework Programme through the project ``OPTIMA-Organization sPecific Threat Intelligence Mining and sharing"~(101063107), funded by the European Union. Views and opinions expressed are however those of the authors only and do not necessarily reflect those of the European Union or the Italian MUR. Neither the European Union nor the Italian MUR can be held responsible for them.

{\em (ii)} The project SERICS (PE00000014) under the NRRP MUR program funded by the EU - NGEU. Views and opinions expressed are however those of the authors only and do not necessarily reflect those of the European Union or the Italian MUR. Neither the European Union nor the Italian MUR can be held responsible for them.
\bibliographystyle{plain}
\bibliography{reference}

\begin{thebibliography}{10}

\bibitem{abdin2024phi}
Marah Abdin, Jyoti Aneja, Hany Awadalla, Ahmed Awadallah, Ammar~Ahmad Awan, Nguyen Bach, Amit Bahree, Arash Bakhtiari, Jianmin Bao, Harkirat Behl, et~al.
\newblock Phi-3 technical report: A highly capable language model locally on your phone.
\newblock {\em arXiv preprint arXiv:2404.14219}, 2024.

\bibitem{al2024machine}
Abdullah~M Al~Alawi, Halima~H Al~Shuaili, Khalid Al-Naamani, Zakariya Al~Naamani, and Said~A Al-Busafi.
\newblock A machine learning-based mortality prediction model for patients with chronic hepatitis c infection: An exploratory study.
\newblock {\em Journal of Clinical Medicine}, 13(10):2939, 2024.

\bibitem{arora2019permpair}
Anshul Arora, Sateesh~K Peddoju, and Mauro Conti.
\newblock Permpair: Android malware detection using permission pairs.
\newblock {\em IEEE Transactions on Information Forensics and Security}, 15:1968--1982, 2019.

\bibitem{bakour2021visdroid}
Khaled Bakour and Halil~Murat {\"U}nver.
\newblock Visdroid: Android malware classification based on local and global image features, bag of visual words and machine learning techniques.
\newblock {\em Neural Computing and Applications}, 33(8):3133--3153, 2021.

\bibitem{bayer2024cysecbert}
Markus Bayer, Philipp Kuehn, Ramin Shanehsaz, and Christian Reuter.
\newblock Cysecbert: A domain-adapted language model for the cybersecurity domain.
\newblock {\em ACM Transactions on Privacy and Security}, 27(2):1--20, 2024.

\bibitem{bohlender2020extreme}
Simon Bohlender, Eneldo Loza~Menc{\'\i}a, and Moritz Kulessa.
\newblock Extreme gradient boosted multi-label trees for dynamic classifier chains.
\newblock In {\em Discovery Science: 23rd International Conference, DS 2020, Thessaloniki, Greece, October 19--21, 2020, Proceedings 23}, pages 471--485. Springer, 2020.

\bibitem{cerri2014hierarchical}
Ricardo Cerri, Rodrigo~C Barros, and Andr{\'e}~CPLF De~Carvalho.
\newblock Hierarchical multi-label classification using local neural networks.
\newblock {\em Journal of Computer and System Sciences}, 80(1):39--56, 2014.

\bibitem{chang2024survey}
Yupeng Chang, Xu~Wang, Jindong Wang, Yuan Wu, Linyi Yang, Kaijie Zhu, Hao Chen, Xiaoyuan Yi, Cunxiang Wang, Yidong Wang, et~al.
\newblock A survey on evaluation of large language models.
\newblock {\em ACM transactions on intelligent systems and technology}, 15(3):1--45, 2024.

\bibitem{charte2015mlsmote}
Francisco Charte, Antonio~J Rivera, Mar{\'\i}a~J del Jesus, and Francisco Herrera.
\newblock Mlsmote: Approaching imbalanced multilabel learning through synthetic instance generation.
\newblock {\em Knowledge-Based Systems}, 89:385--397, 2015.

\bibitem{chawla2002smote}
Nitesh~V Chawla, Kevin~W Bowyer, Lawrence~O Hall, and W~Philip Kegelmeyer.
\newblock Smote: synthetic minority over-sampling technique.
\newblock {\em Journal of artificial intelligence research}, 16:321--357, 2002.

\bibitem{chen2023unleashing}
Banghao Chen, Zhaofeng Zhang, Nicolas Langren{\'e}, and Shengxin Zhu.
\newblock Unleashing the potential of prompt engineering in large language models: a comprehensive review.
\newblock {\em arXiv preprint arXiv:2310.14735}, 2023.

\bibitem{chen2025aecr}
Minghao Chen, Kaijie Zhu, Bin Lu, Ding Li, Qingjun Yuan, and Yuefei Zhu.
\newblock Aecr: Automatic attack technique intelligence extraction based on fine-tuned large language model.
\newblock {\em Computers \& Security}, 150:104213, 2025.

\bibitem{dettmers2024qlora}
Tim Dettmers, Artidoro Pagnoni, Ari Holtzman, and Luke Zettlemoyer.
\newblock Qlora: Efficient finetuning of quantized llms.
\newblock {\em Advances in Neural Information Processing Systems}, 36, 2024.

\bibitem{fairbanks2021identifying}
Jeffrey Fairbanks, Andres Orbe, Christine Patterson, Janet Layne, Edoardo Serra, and Marion Scheepers.
\newblock Identifying att\&ck tactics in android malware control flow graph through graph representation learning and interpretability.
\newblock In {\em 2021 IEEE International Conference on Big Data (Big Data)}, pages 5602--5608. IEEE, 2021.

\bibitem{fieblinger2024actionable}
Romy Fieblinger, Md~Tanvirul Alam, and Nidhi Rastogi.
\newblock Actionable cyber threat intelligence using knowledge graphs and large language models.
\newblock In {\em 2024 IEEE European Symposium on Security and Privacy Workshops (EuroS\&PW)}, pages 100--111. IEEE, 2024.

\bibitem{goyal2010literature}
Priyanka Goyal, Sahil Batra, and Ajit Singh.
\newblock A literature review of security attack in mobile ad-hoc networks.
\newblock {\em International Journal of Computer Applications}, 9(12):11--15, 2010.

\bibitem{hu2024llm}
Yuelin Hu, Futai Zou, Jiajia Han, Xin Sun, and Yilei Wang.
\newblock Llm-tikg: Threat intelligence knowledge graph construction utilizing large language model.
\newblock {\em Computers \& Security}, 145:103999, 2024.

\bibitem{husari2017ttpdrill}
Ghaith Husari, Ehab Al-Shaer, Mohiuddin Ahmed, Bill Chu, and Xi~Niu.
\newblock Ttpdrill: Automatic and accurate extraction of threat actions from unstructured text of cti sources.
\newblock In {\em Proceedings of the 33rd annual computer security applications conference}, pages 103--115, 2017.

\bibitem{islam2023android}
Rejwana Islam, Moinul~Islam Sayed, Sajal Saha, Mohammad~Jamal Hossain, and Md~Abdul Masud.
\newblock Android malware classification using optimum feature selection and ensemble machine learning.
\newblock {\em Internet of Things and Cyber-Physical Systems}, 3:100--111, 2023.

\bibitem{kim2022comparative}
Heejung Kim and Hwankuk Kim.
\newblock Comparative experiment on ttp classification with class imbalance using oversampling from cti dataset.
\newblock {\em Security and Communication Networks}, 2022(1):5021125, 2022.

\bibitem{kim2018multimodal}
TaeGuen Kim, BooJoong Kang, Mina Rho, Sakir Sezer, and Eul~Gyu Im.
\newblock A multimodal deep learning method for android malware detection using various features.
\newblock {\em IEEE Transactions on Information Forensics and Security}, 14(3):773--788, 2018.

\bibitem{kumar2024prompt}
Neha~Mohan Kumar, Fahmida~Tasnim Lisa, and Sheikh~Rabiul Islam.
\newblock Prompt chaining-assisted malware detection: A hybrid approach utilizing fine-tuned llms and domain knowledge-enriched cybersecurity knowledge graphs.
\newblock In {\em 2024 IEEE International Conference on Big Data (BigData)}, pages 1672--1677. IEEE, 2024.

\bibitem{legoy2020automated}
Valentine Legoy, Marco Caselli, Christin Seifert, and Andreas Peter.
\newblock Automated retrieval of att\&ck tactics and techniques for cyber threat reports.
\newblock {\em arXiv preprint arXiv:2004.14322}, 2020.

\bibitem{lewis2020retrieval}
Patrick Lewis, Ethan Perez, Aleksandra Piktus, Fabio Petroni, Vladimir Karpukhin, Naman Goyal, Heinrich K{\"u}ttler, Mike Lewis, Wen-tau Yih, Tim Rockt{\"a}schel, et~al.
\newblock Retrieval-augmented generation for knowledge-intensive nlp tasks.
\newblock {\em Advances in neural information processing systems}, 33:9459--9474, 2020.

\bibitem{qiu2020survey}
Junyang Qiu, Jun Zhang, Wei Luo, Lei Pan, Surya Nepal, and Yang Xiang.
\newblock A survey of android malware detection with deep neural models.
\newblock {\em ACM Computing Surveys (CSUR)}, 53(6):1--36, 2020.

\bibitem{saracino2016madam}
Andrea Saracino, Daniele Sgandurra, Gianluca Dini, and Fabio Martinelli.
\newblock Madam: Effective and efficient behavior-based android malware detection and prevention.
\newblock {\em IEEE Transactions on Dependable and Secure Computing}, 15(1):83--97, 2016.

\bibitem{shafee2024evaluation}
Samaneh Shafee, Alysson Bessani, and Pedro~M Ferreira.
\newblock Evaluation of llm-based chatbots for osint-based cyber threat awareness.
\newblock {\em Expert Systems with Applications}, page 125509, 2024.

\bibitem{sharma2023radar}
Yashovardhan Sharma, Simon Birnbach, and Ivan Martinovic.
\newblock Radar: A ttp-based extensible, explainable, and effective system for network traffic analysis and malware detection.
\newblock In {\em Proceedings of the 2023 European Interdisciplinary Cybersecurity Conference}, pages 159--166, 2023.

\bibitem{sharma2023ttp}
Yashovardhan Sharma, Eleonora Giunchiglia, Simon Birnbach, and Ivan Martinovic.
\newblock To ttp or not to ttp?: Exploiting ttps to improve ml-based malware detection.
\newblock In {\em 2023 IEEE International Conference on Cyber Security and Resilience (CSR)}, pages 8--15. IEEE, 2023.

\bibitem{shen2024ghgdroid}
Lina Shen, Mengqi Fang, and Jian Xu.
\newblock Ghgdroid: Global heterogeneous graph-based android malware detection.
\newblock {\em Computers \& Security}, 141:103846, 2024.

\bibitem{shin2023exploiting}
Chanho Shin, Insup Lee, and Changhee Choi.
\newblock Exploiting ttp co-occurrence via glove-based embedding with mitre att\&ck framework.
\newblock {\em IEEE Access}, 2023.

\bibitem{tsoumakas2010mining}
Grigorios Tsoumakas, Ioannis Katakis, and Ioannis Vlahavas.
\newblock Mining multi-label data.
\newblock {\em Data mining and knowledge discovery handbook}, pages 667--685, 2010.

\bibitem{van2022digital}
Ziboud Van~Veldhoven and Jan Vanthienen.
\newblock Digital transformation as an interaction-driven perspective between business, society, and technology.
\newblock {\em Electronic markets}, 32(2):629--644, 2022.

\bibitem{vens2008decision}
Celine Vens, Jan Struyf, Leander Schietgat, Sa{\v{s}}o D{\v{z}}eroski, and Hendrik Blockeel.
\newblock Decision trees for hierarchical multi-label classification.
\newblock {\em Machine learning}, 73:185--214, 2008.

\bibitem{vinayakumar2017deep}
R~Vinayakumar, KP~Soman, and Prabaharan Poornachandran.
\newblock Deep android malware detection and classification.
\newblock In {\em 2017 International conference on advances in computing, communications and informatics (ICACCI)}, pages 1677--1683. IEEE, 2017.

\bibitem{wu2021android}
Bozhi Wu, Sen Chen, Cuiyun Gao, Lingling Fan, Yang Liu, Weiping Wen, and Michael~R Lyu.
\newblock Why an android app is classified as malware: Toward malware classification interpretation.
\newblock {\em ACM Transactions on Software Engineering and Methodology (TOSEM)}, 30(2):1--29, 2021.

\bibitem{wu2019multi}
Xin Wu, Yuchen Gao, and Dian Jiao.
\newblock Multi-label classification based on random forest algorithm for non-intrusive load monitoring system.
\newblock {\em Processes}, 7(6):337, 2019.

\bibitem{yan2018survey}
Ping Yan and Zheng Yan.
\newblock A survey on dynamic mobile malware detection.
\newblock {\em Software Quality Journal}, 26(3):891--919, 2018.

\bibitem{yang2024harnessing}
Jingfeng Yang, Hongye Jin, Ruixiang Tang, Xiaotian Han, Qizhang Feng, Haoming Jiang, Shaochen Zhong, Bing Yin, and Xia Hu.
\newblock Harnessing the power of llms in practice: A survey on chatgpt and beyond.
\newblock {\em ACM Transactions on Knowledge Discovery from Data}, 18(6):1--32, 2024.

\bibitem{yoga975hybrid}
Castro~A Yoga, Anthony~J Rodrigues, and Silvance~O Abeka.
\newblock Hybrid machine learning approach for attack classification and clustering in network security.
\newblock {\em International Journal of Computer Applications}, 975:8887, 2023.

\bibitem{zhang2025attackg+}
Yongheng Zhang, Tingwen Du, Yunshan Ma, Xiang Wang, Yi~Xie, Guozheng Yang, Yuliang Lu, and Ee-Chien Chang.
\newblock Attackg+: Boosting attack graph construction with large language models.
\newblock {\em Computers \& Security}, 150:104220, 2025.

\bibitem{zulkefli2020sentient}
Zakiah Zulkefli and Manmeet~Mahinderjit Singh.
\newblock Sentient-based access control model: a mitigation technique for advanced persistent threats in smartphones.
\newblock {\em Journal of Information Security and Applications}, 51:102431, 2020.

\end{thebibliography}
\clearpage

\end{document}